\documentclass[preprint,aps,amsmath,showkeys, nofootinbib,
superscriptaddress]{revtex4} 
\usepackage{ulem}
\usepackage{graphicx} 
\usepackage{color}
 
\begin{document} 

\title{\textbf{Single- and two-nucleon antikaon absorption in nuclear matter with chiral meson-baryon interactions}} 

\author{J.~Hrt\'{a}nkov\'{a}} 
\email{hrtankova@ujf.cas.cz} 
\affiliation{Nuclear Physics Institute of the Czech Academy of Sciences, 25068 \v{R}e\v{z}, Czech Republic} 

\author{\`{A}.~Ramos} 
\affiliation{Departament de Física Quàntica i Astrofísica and Institut de Ciencies del Cosmos, Universitat de Barcelona, Barcelona, Spain} 

\date{\today} 

\begin{abstract} 
We developed a microscopic model for antikaon absorption on two nucleons in nuclear matter. The absorption is described within a meson-exchange picture and the primary $K^-N$ interaction strength is derived from state-of-the-art chiral coupled channel meson-baryon interaction models. We took into account the medium modification of the $K^-N$ scattering amplitudes due to the Pauli correlations. We derived the $K^-NN$ as well as $K^-N$ optical potentials as functions of nuclear matter density including the real part of the $K^-NN$ potential. We calculated the $K^-$ single- and two-nucleon absorption fractions and branching ratios for various mesonic and non-mesonic channels. We confirmed the crucial role of in-medium effects in our calculations. Our results are in very good agreement with available experimental data from old bubble chamber experiments as well as with the latest results from the AMADEUS collaboration.
\end{abstract}


\maketitle 

\section{Introduction}
\label{Intro}

The absorption of $K^-$ on two or more nucleons represents about 20\% of all $K^-$ absorption in the surface region of atomic nuclei. The multi-nucleon absorption ratios were first measured in the 1960's and 1970's in bubble chamber experiments \cite{bubble1,bubble2,bubble3, katzPRD70}. The multi-nucleon absorption fraction measured for $K^-$ capture on a mixture of C, F, and Br was found to be $0.26\pm0.03$~\cite{bubble1},  while Ref.~\cite{bubble2} quotes $0.28\pm0.03$ for capture on Ne and Ref.~\cite{bubble3} lists the value $0.19\pm0.03$ for capture on C. Katz et al.~\cite{katzPRD70} measured the $K^-$ two-nucleon absorption fractions for all possible final states on $^4$He and Veirs and Burnstein \cite{veirsPRD70} measured the $K^-$ two-nucleon absorption fractions on the deuteron. A detailed kinematic analysis of the reaction $K^- + ^{4}\text{He}\rightarrow \Lambda + d+ n$ was given in Ref.~\cite{bubble4}, together with a branching ratio for this process of $0.035\pm0.002$. In the past decade, experiment E549 at KEK measured $K^-$ three- and four-nucleon absorption fractions on $^{4}$He for channels with a $\Lambda$ hyperon in the final state \cite{KEK1}. The FINUDA collaboration analyzed the $\Sigma^- p$ emission rate in reactions of low-energy antikaons with light nuclei ($6\leq A \leq16$) \cite{FINUDA}.
Very recently, the AMADEUS collaboration measured the $K^-$ two-nucleon branching ratios with $\Lambda p$ and $\Sigma^0 p$ in the final state for low-energy antikaons absorbed by a carbon target~\cite{amadeus16, amadeus19}. The ratio of branching ratios $R={\rm BR}(K^-pp\rightarrow \Lambda p)/{\rm BR}(K^-pp\rightarrow \Sigma^0 p)$ was found to be around $0.7$ \cite{amadeus19}. All these measurements provided valuable and detailed information about the $K^-$ multi-nucleon absorption processes.

The $K^-N$ interaction is known to be attractive in the medium due to the subthreshold $I=0$ resonance $\Lambda(1405)$, which couples strongly to the $\pi \Sigma$ channel giving rise to a sizable $K^-$ absorption. The theoretical description of the $K^-N$ interaction is currently provided by the chiral coupled channel meson-baryon interactions models \cite{pnlo, kmnlo, bcn, m, b} in which the $\Lambda(1405)$ is generated dynamically. Parameters of these models are fitted to available low-energy $K^-N$ observables \cite{kp_ratios1, kp_ratios2, kp_crosssection1, kp_crosssection2, kp_crosssection3, SIDDHARTA}. On the other hand, the interaction of antikaons with two and more nucleons lacks a solid theoretical description.

A very important source of information about the $K^-$-nucleus interaction is provided by kaonic atom experiments, in which the low-energy $K^-$ annihilates in the surface region of a nucleus, thereby probing the $K^-$-nucleus potential at low nuclear densities close to threshold. The latest analysis of kaonic atom data by Friedman and Gal \cite{fgNPA17} showed that $K^-$ optical potentials based on the $K^-N$ scattering amplitudes derived from state-of-the-art chiral models fail in general to describe the data unless a purely phenomenological term representing the $K^-$ multi-nucleon interaction is added. Moreover, after applying an extra constraint to reproduce the $K^-$ single-nucleon absorption fractions from bubble chamber experiments \cite{bubble1, bubble2, bubble3}, only three models, namely the Prague model (P) \cite{pnlo}, the Kyoto-Munich model (KM) \cite{kmnlo}, and the Barcelona model (BCN) \cite{bcn}, were found acceptable. The $K^-$ multi-nucleon interaction is thus an inseparable part of any realistic description of the $K^-$-nucleus interaction.

The attractive nature of the $K^-N$ interaction led to conjectures about the existence of $K^-$ bound states. The first ever observation of the $K^-pp$ bound state was reported recently by the J-PARC E15 Collaboration \cite{Sada:2016nkb,Ajimura:2018iyx} in reactions employing in-flight antikaons on a $^3$He target. It has also a solid theoretical support from the study of Ref.~\cite{Sekihara:2016vyd}.
However, the $K^-$ multi-nucleon absorption may have serious implications for existence of the $K^-$-nuclear states, particularly in heavier systems. The $K^-$ single-nucleon optical potential based on the P and KM models, supplemented by a phenomenological multi-nucleon potential fitted to reproduce kaonic atom data, was applied in the calculations of $K^-$ quasibound states in nuclei with $A\geq6$ \cite{hmPLB_PRC}. The multi-nucleon absorption potential was found to have a significant contribution to the total $K^-$-nuclear absorptive potential. The widths of $K^-$-nuclear quasibound states resulted to be up to one order of magnitude larger than the corresponding binding energies. It is to be noted that kaonic atom data can probe reliably the $K^-$ potential only up to $\sim 50\%$ of normal nuclear density $\rho_0$. Therefore, the evaluation of $K^-$ multi-nucleon potential around $\rho_0$ explored by the $K^-$-nuclear quasibound states is a mere extrapolation or analytical continuation of the empirical formula.

It is then clear that, to enrich our knowledge about the $K^-$ absorption on two or more nucleons at any density, a sound theoretical microscopic approach is needed. It should be connected to the $K^-N$ chiral interaction models in order to provide a unified description of the $K^-$ single- and multi-nucleon potential. The theoretical description of the $K^-NN$ absorption via meson rescattering process was already proposed in 1989 by Onaga et al. \cite{onagaPTP89}. Sekihara et al. \cite{sjPRC79} connected the two-nucleon antikaon absorption potential to the modern chiral $\bar{K}N$ interaction in a first exploratory study of the non-mesonic absorption of the $\Lambda(1405)$ in nuclear matter via a one-meson exchange mechanism. Transition probabilities to $\Lambda N$ and $\Sigma N$ final states were calculated, showing that the ratio $\Gamma_{\Lambda N}/ \Gamma_{\Sigma N}=1.2$ is independent of the nuclear density when employing the couplings of the $\Lambda(1405)$ to $\bar{K}N$ and $\pi \Sigma$ states from a chiral unitary meson-baryon interaction model. In a subsequent work~\cite{sjPRC12}, the authors developed a microscopic model for the $K^-NN$ absorption in nuclear matter using the free-space $K^-N$ scattering amplitudes derived from a chiral meson-baryon interaction model. They described the $K^-NN$ absorption within the meson-exchange picture and calculated the imaginary part of the $K^-NN$ self-energy.  

In this paper, we present a microscopic model for the $K^-NN$ absorption in symmetric nuclear matter. It is motivated by the approach of Nagahiro et al. \cite{nagahiroPLB12} where a method for obtaining the meson-nucleon-nucleon self-energy within a meson-exchange picture was developed and applied to calculate the $\eta'NN$ optical potential in nuclear matter. We extend the approach to incorporate also the exchange terms which are non-negligible in the absorption of low-energy antikaons studied in the present work. 
In the formalism employed here, the absorption of $K^-$ on two nucleons is modeled within a meson-exchange picture and the $K^-NN$ optical potential is derived from the corresponding $K^-$ self-energy. The primary $K^-N$ interaction strength is provided by the $K^-N$ scattering amplitudes derived from the P and BCN chiral models. Unlike Ref.~\cite{sjPRC12}, we take into account the Pauli correlations in the medium for the $K^-N$ scattering amplitudes since they are essential for in-medium kinematics. Moreover, we present results for the real part of the $K^-NN$ optical potential derived within the microscopic model for the first time. 

We calculate branching ratios for the $K^-$ single- and two-nucleon absorption channels and compare our findings with old bubble chamber data as well as with the latest measurements from the AMADEUS collaboration \cite{amadeus19}. In general, our results reproduce the experimental data remarkably well, especially when the in-medium effects are taken into account.

The paper is organized as follows. In Section \ref{formalism}, we give a brief description of the formalism used to derive the $K^-N$ and $K^-NN$ optical potentials and basic information about the chiral coupled channel meson-baryon interaction models used in our approach. In Section \ref{results}, we present the real and imaginary parts of the $K^-N$ and $K^-NN$ optical potentials calculated within our model and compare the branching ratios for mesonic and non-mesonic $K^-$ decay channels with data. Finally, we summarize our findings in Section \ref{conclusions}. Detailed derivation of the $K^-N$ and $K^-NN$ optical potentials is given in Appendices~\ref{appendix:a} and \ref{appendix:b}, respectively.

\section{Formalism}
\label{formalism}

In this section we give a brief description of the formalism used to derive the $K^-N$ and $K^-NN$ absorption potentials in symmetric nuclear matter. More details can be found in Appendices~\ref{appendix:a} and \ref{appendix:b}. The $K^-N \to M B$ amplitudes appearing in the one-nucleon and two-nucleon absorption diagrams are obtained from chiral coupled channel meson-baryon interaction models \cite{bcn,pnlo}. A derivation of a $K^-$ absorption potential using chiral two-body scattering amplitudes was applied before by Sekihara et al.~\cite{sjPRC12}. In the present work we consider in-medium $K^-N \to M B$ amplitudes modified by Pauli blocking effects.

\subsection{Single-nucleon $K^-$ absorption in nuclear matter}

 \begin{figure}[ht!]
\includegraphics[width=0.2\textwidth]{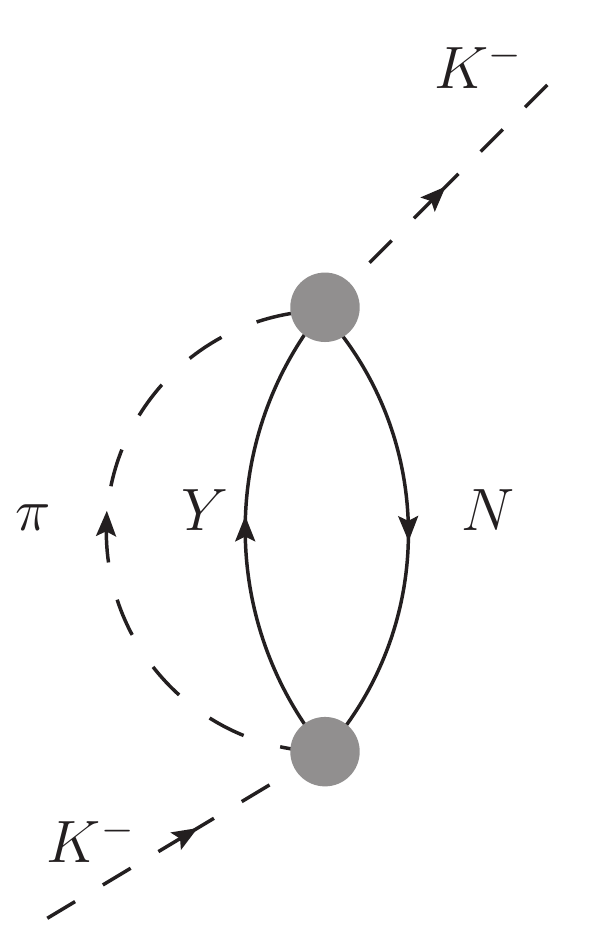}
\caption{\label{fig:one_loop_diagrams} Feynman diagram for mesonic $K^-$ absorption in nuclear matter. The shaded circles denote the $K^-N\rightarrow \pi Y,~(Y=\Lambda,~\Sigma)$ t-matrices derived from a chiral coupled channel meson-baryon interaction model.}
\end{figure}
Let us start with the 1$N$-absorption mechanism. The interaction of a kaon $K^-$ with a single nucleon $N=p,~n$ in nuclear matter is depicted by the Feynman diagram of Fig.~\ref{fig:one_loop_diagrams}. The algebraic expression of the  $K^-N$ self-energy is given by Eq.~\eqref{eq:PiKN1} in Appendix~\ref{appendix:a}, where all the details of the calculation are thoroughly discussed. Here, we only note that the imaginary part of the corresponding $K^-N$ optical potential, given by Eq.~\eqref{eq:imVknpiY}, is obtained from the sum of the contributions from different annihilation channels listed in Table~\ref{tab:channels}
\begin{equation}\label{eq:imVkn}
 \text{Im}V_{K^-N}=\sum_{\rm channels} \text{Im}V_{K^-N\rightarrow \pi Y}~.
\end{equation}
We have checked that the sum in Eq.~\eqref{eq:imVkn} gives the same total one-nucleon absorption width obtained directly from a $t \rho$-type expression. Relatedly, the real part of the single-nucleon potential, presented in Section~\ref{results}, is obtained from the $t\rho$ expression as well.

\begin{table}[h!]
\caption{All considered channels for mesonic and non-mesonic $K^-$ absorption in nuclear matter. }
 \begin{tabular}{cl||cl} \label{tab:channels}
  $K^-N$ & $\rightarrow \pi Y$ & $K^-N_1N_2$  &$\rightarrow YN$ \\ \hline \hline
   $K^-p$& $\rightarrow \pi^0 \Lambda$ & $K^-pp$ & $\rightarrow \Lambda p$ \\
   & $\rightarrow \pi^0 \Sigma^0$ & & $\rightarrow \Sigma^0 p$ \\
   & $\rightarrow \pi^+ \Sigma^-$ & & $\rightarrow \Sigma^+ n$  \\
  & $\rightarrow \pi^- \Sigma^+$ & $K^-pn(np)$ & $\rightarrow \Lambda n$ \\ 
  $K^-n$ & $\rightarrow \pi^- \Lambda$  & & $\rightarrow \Sigma^0 n$ \\ 
  & $\rightarrow \pi^- \Sigma^0$ & & $\rightarrow \Sigma^- p$ \\
  & $\rightarrow \pi^0 \Sigma^-$ & $K^-nn$ & $\rightarrow \Sigma^- n $ \\ \hline
 \end{tabular}

\end{table}

\subsection{Two-nucleon $K^-$ absorption in nuclear matter} 

\begin{figure}[t!]
\includegraphics[width=0.7\textwidth]{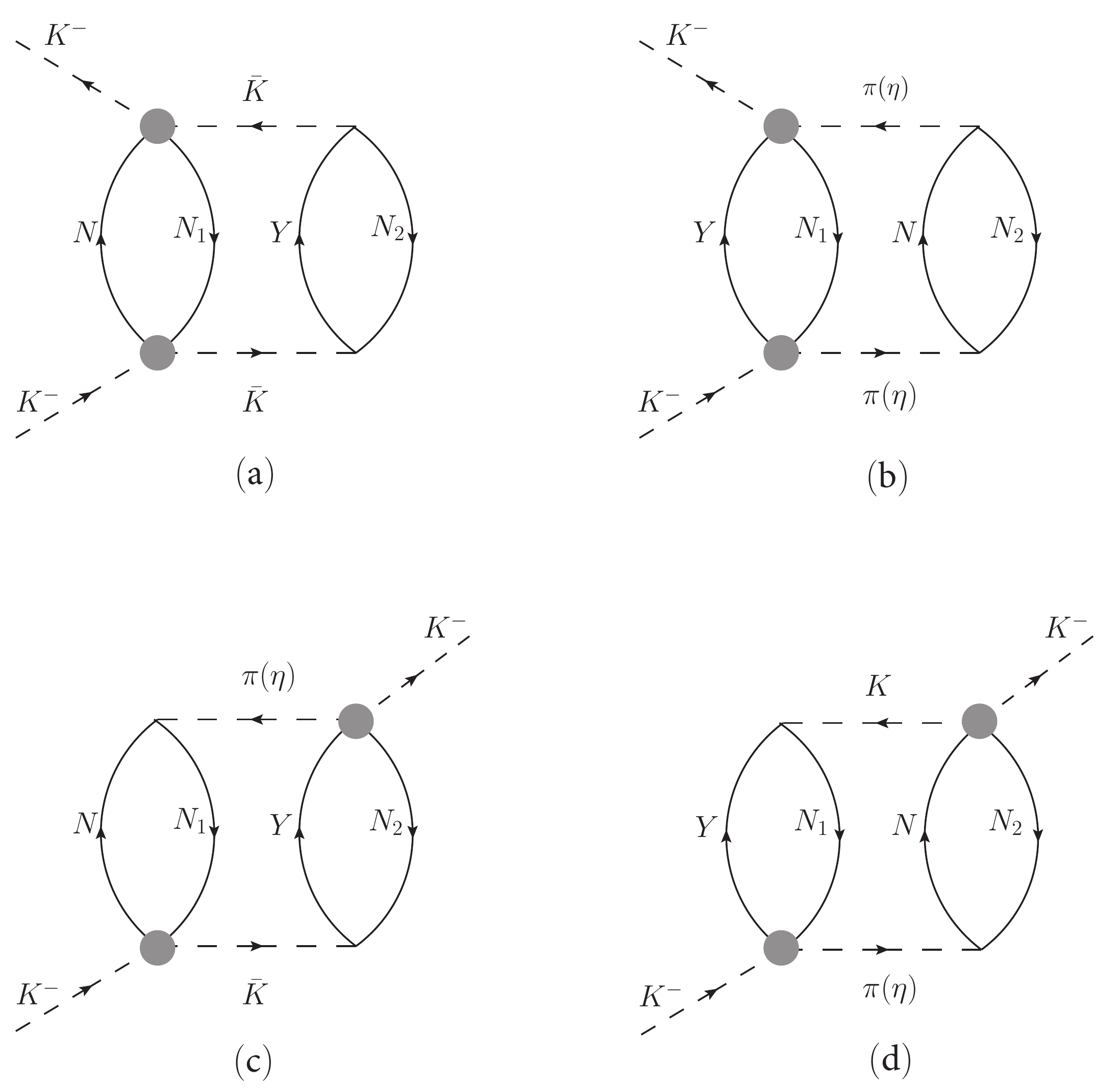}
\caption{\label{fig:direct_diagrams} Two-fermion-loop Feynman diagrams for non-mesonic $K^-$ absorption on two nucleons $N_1,~N_2$ in nuclear matter. The shaded circles denote the $K^-N$ t-matrices derived from a chiral coupled channel meson-baryon interaction model. }
\end{figure}

\begin{figure}[t!]
\includegraphics[width=0.7\textwidth]{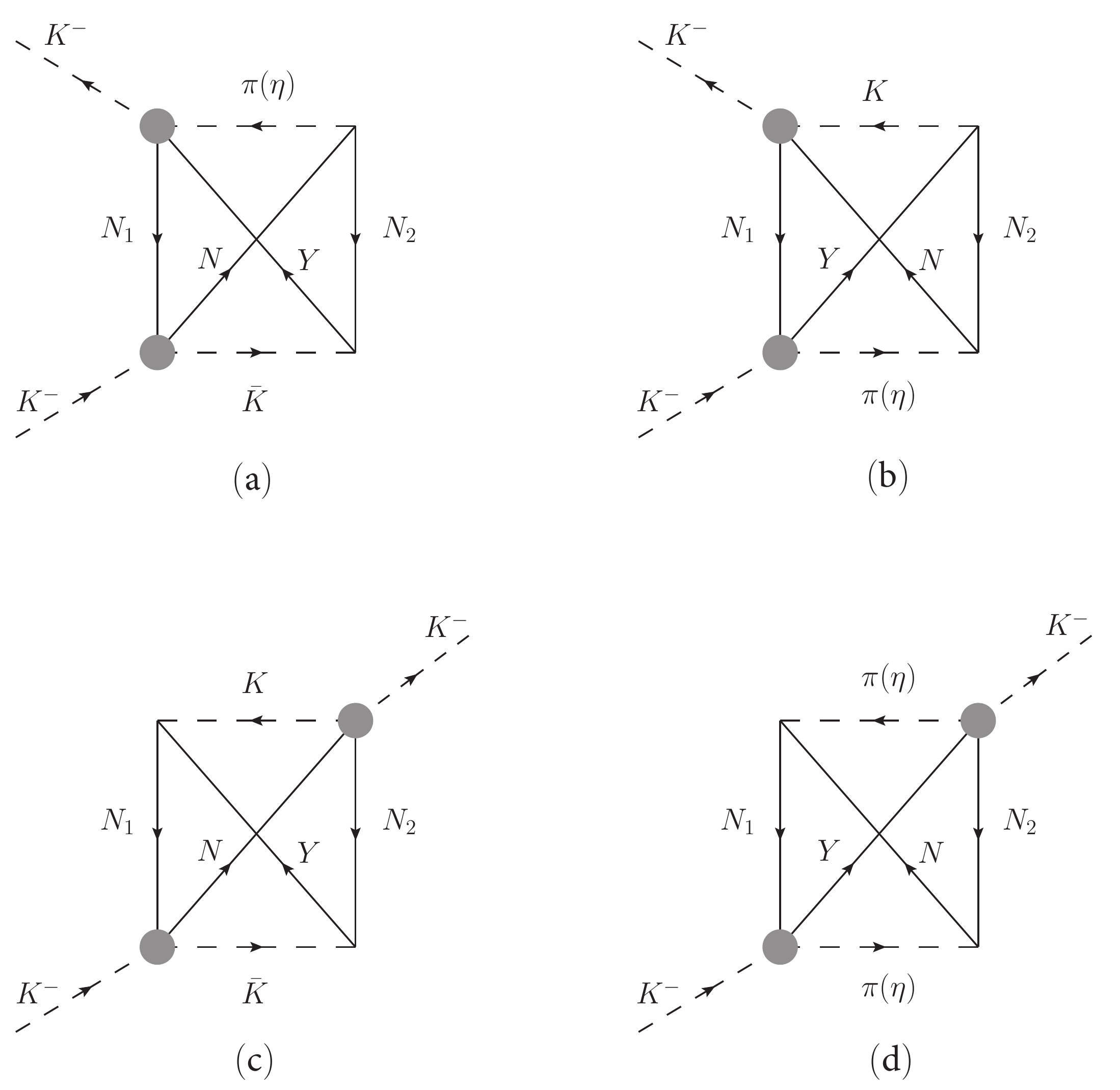}
\caption{\label{fig:crossed_diagrams} One-fermion-loop Feynman diagrams for non-mesonic $K^-$ absorption on two nucleons $N_1,~N_2$ in nuclear matter. The shaded circles denote the $K^-N$ t-matrices derived from a chiral coupled channel meson-baryon interaction model.}
\end{figure}

The absorption of $K^-$ by two nucleons in nuclear matter is described within a meson-exchange picture. Our formalism for 2$N$-absorption follows closely the approach of Nagahiro et al.~\cite{nagahiroPLB12} used to derive the $\eta^{\prime}NN$ optical potential, except that in the present work we also consider the non-negligible effect of the exchange terms. The Feynman diagrams representing the $2N$-absorption process with different intermediate virtual mesons exchanged ($\overline{K},~\pi,~\eta$) are depicted in Figs.~\ref{fig:direct_diagrams} and \ref{fig:crossed_diagrams}. The shaded circles denote the $K^-N$ t-matrices derived from a chiral coupled channel meson-baryon interaction model. We refer to the diagrams shown in Fig.~\ref{fig:direct_diagrams}, which provide the main contribution to the $K^-NN$ self-energy, as `two fermion loop (2FL)'. We refer to the diagrams (a) and (b) in Fig.~\ref{fig:crossed_diagrams} as `one fermion loop of type A (1FLA)' and to the diagrams (c) and (d) as `one fermion loop of type B (1FLB)'. Note that the direct contributions correspond to diagrams 2FL(a) and 2FL(b) in Fig.~\ref{fig:direct_diagrams}, while the remaining diagrams in this figure and those of Fig.~\ref{fig:crossed_diagrams} are obtained from antisymmetrizing the initial $N_1N_2$ system, as well as from exchanging the place of the $N$ and $Y$ baryons in the final state. The considered channels for two-body $K^-$ absorption in nuclear matter are listed in the second column of Table~\ref{tab:channels}. Each channel in that list can proceed via direct and exchange diagrams with the corresponding intermediate mesons. In this way, our approach incorporates the same $2N$-absorption processes as those studied in Ref.~\cite{sjPRC12}.  

The total $K^-NN$ potential is then built as a sum of contributions coming from the 2FL and 1FL diagrams for all considered channels listed in Table~\ref{tab:channels}
\begin{equation}
 V_{K^-NN}=\sum_{\rm channels} V_{K^-NN}^{\rm 2FL} + V_{K^-NN}^{\rm 1FLA} + V_{K^-NN}^{\rm 1FLB}~.
\end{equation}
For illustration, there are 37 2FL-type diagrams, 28 1FLA-type diagrams and 33 1FLB-type diagrams that contribute to the total $K^-NN$ optical potential. The details of the derivation of the 2FL and 1FL $K^-NN$ self-energies and the explicit forms of their respective optical potentials are given in Appendix~\ref{appendix:b}.

\begin{figure}[t!]
\begin{center}
\includegraphics[width=0.75\textwidth]{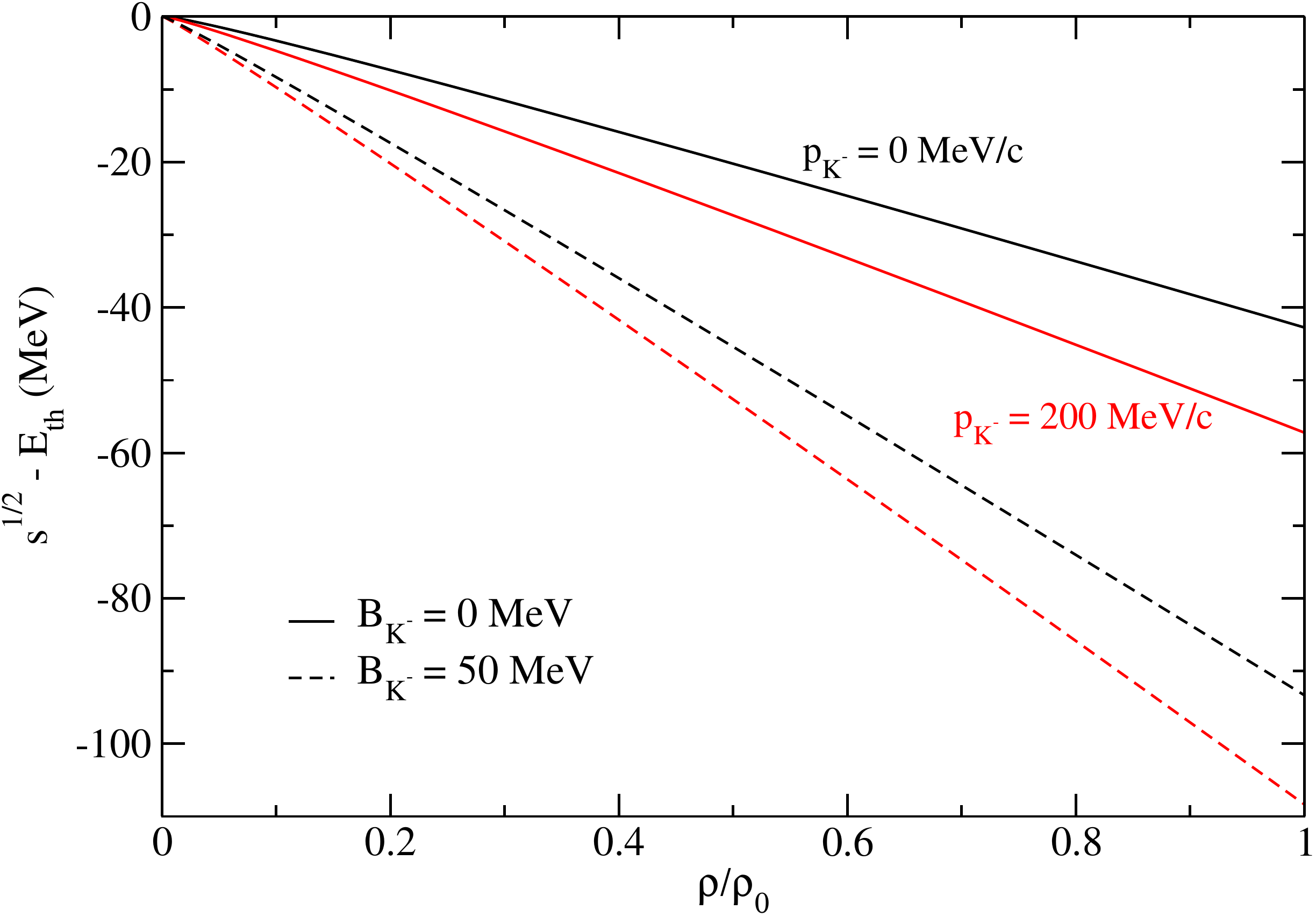}
\end{center}
\caption{The energy shift $\delta \sqrt{s}=\sqrt{s}- E_{\rm th}$ probed in our model as a function of the relative density $\rho/\rho_0$ for two values of the kaon binding energy $B_{K^-}$ and momentum $p_{K^-}$.}
\label{fig:deltaEs}
\end{figure}

The argument of the t-matrices used in the evaluation of the $K^-$ optical potential is the center-of-mass energy $\sqrt{s}$. In the discussion of our results for the $K^-$ optical potential it is useful to know the range of values this parameter may take. As we consider the interaction of an external $K^-$ with a nucleon in nuclear matter, the expression for $\sqrt{s}$ reads
\begin{equation}\label{eq:sqrtS}
 \sqrt{s}=\sqrt{(E_{K^-}+\langle E_N \rangle)^2 - \langle k\rangle^2 -p^2_{K^-}\frac{\rho}{\rho_0}}~,
\end{equation}
where $E_{K^-}=m_{K^-}-B_{K^-}\frac{\rho}{\rho_0}$, $\langle E_N \rangle$ is the average nucleon energy given by Eq.~\eqref{eq:EN}, $\langle k\rangle~=~\sqrt{\frac{3}{5}}\;k_F$ is the average nucleon momentum and $p_{K^-}$ is the kaon momentum (we average over the angles, i.~e. $(\langle k\rangle+\vec{p}_{K^-})^2\rightarrow \langle k\rangle^2 + p^2_{K^-}$). Here, $m_{K^-}$ is the antikaon mass, $B_{K^-}$ denotes the $K^-$ binding energy at saturation density $\rho_0$, $\rho$ is the nuclear matter density and $k_F$ is the corresponding Fermi momentum. We multiply the kaon momentum by a square root of relative density, i.~e. $p_{K^-} \rightarrow p_{K^-}(\frac{\rho}{\rho_0})^{1/2}$ in order to maintain the low-density limit in $\sqrt{s}$, i.~e. $\sqrt{s} \rightarrow m_{K^-}+m_N$ as $\rho \rightarrow 0$.

In Fig.~\ref{fig:deltaEs} we present the energy shift $\delta \sqrt{s} = \sqrt{s} - E_{\rm th}$,  with $E_{\rm th}=m_{K^-}+m_N$, as a function of the relative density $\rho/\rho_0$ for two values of the kaon binding energy $B_{K^-}=0$~MeV and $50$~MeV and two different kaon momenta $p_{K^-}= 0$~MeV/c and $200$~MeV/c\footnote{This value corresponds to a momentum of a kaon bound by $50$~MeV in a potential $V_{K^-}\sim~-~80$~MeV at~$\rho_0$.} at $\rho_0$. In the case of kaon at rest, $p_{K^-}=0$~MeV/c (black), we probe energies down to $\sim 40$~MeV below threshold at saturation density for $B_{K^-}=0$~MeV. If the value $B_{K^-}$ increases to 50~MeV we probe lower energies, approaching the $\pi \Sigma$ threshold at saturation density. If we assign the kaon a momentum $p_{K^-}=200$~MeV/c (red) at saturation density, the absolute value of the energy shift increases by about 16~MeV at $\rho_0$ for both values of $B_{K^-}$.

\subsection{${\overline K}N$ interaction models}
\label{theory}

\begin{figure}[t!]
\begin{center}
\includegraphics[width=0.48\textwidth]{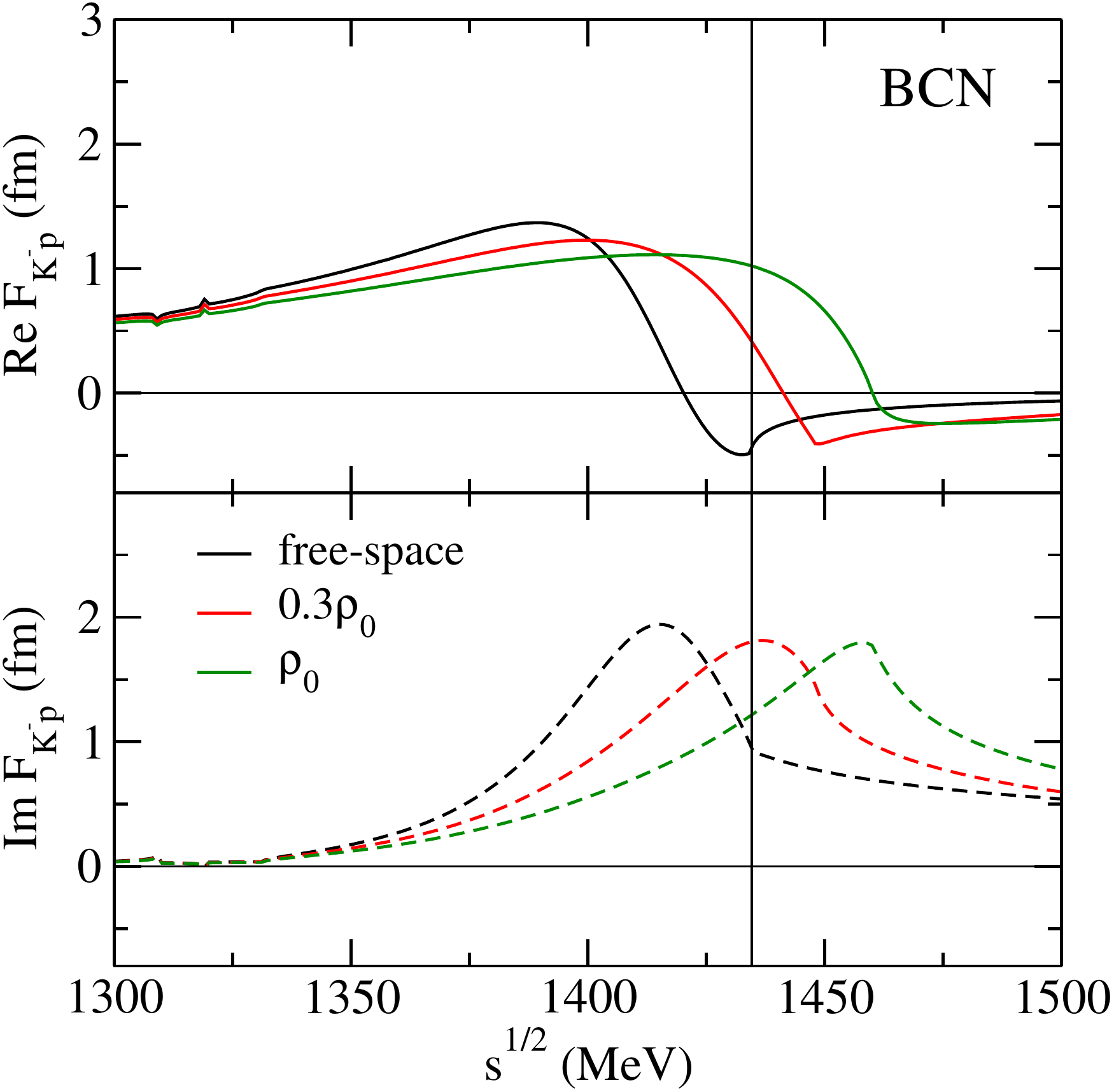} \hspace{10pt}
\includegraphics[width=0.48\textwidth]{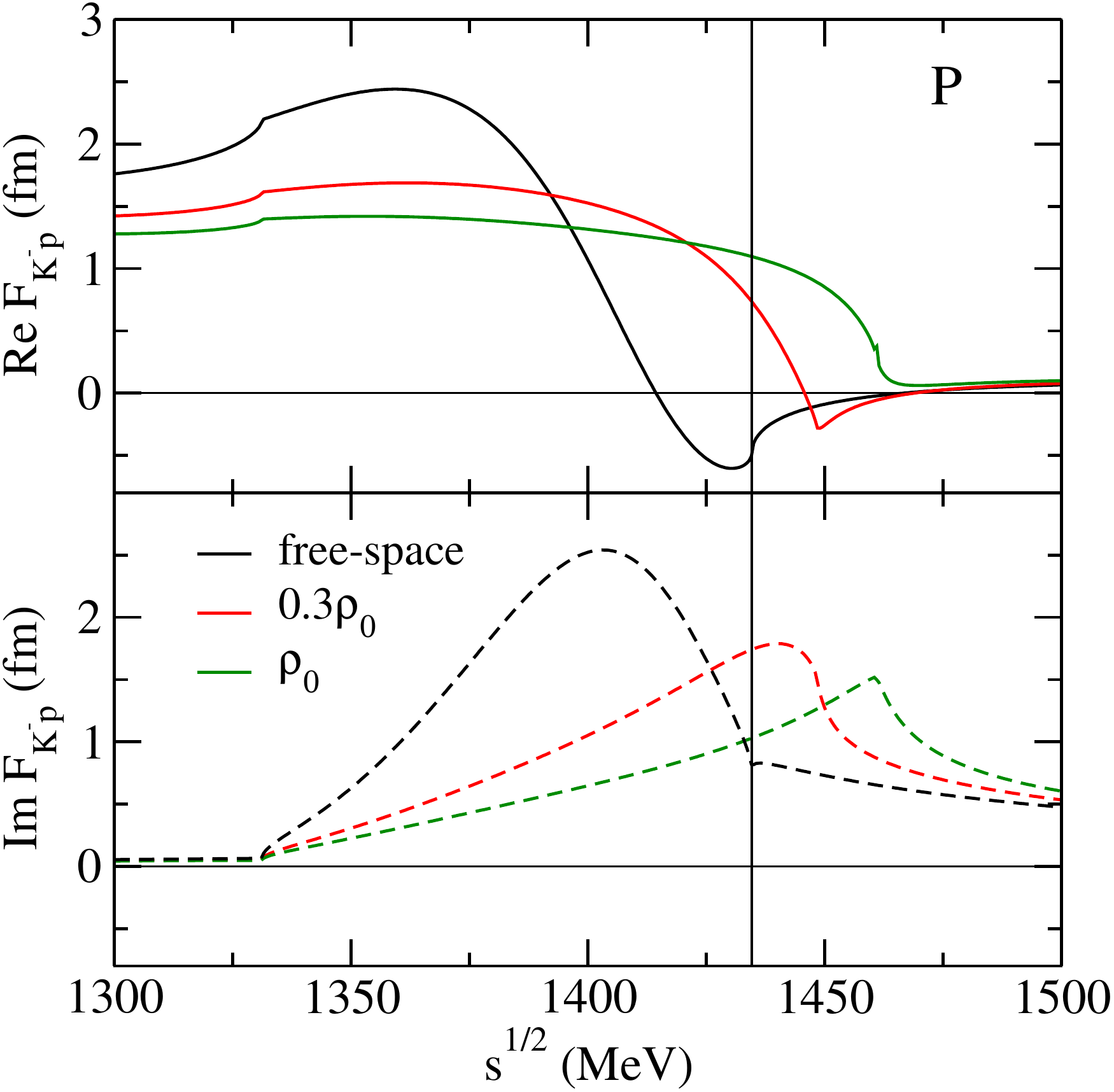} \\[2ex]
\includegraphics[width=0.48\textwidth]{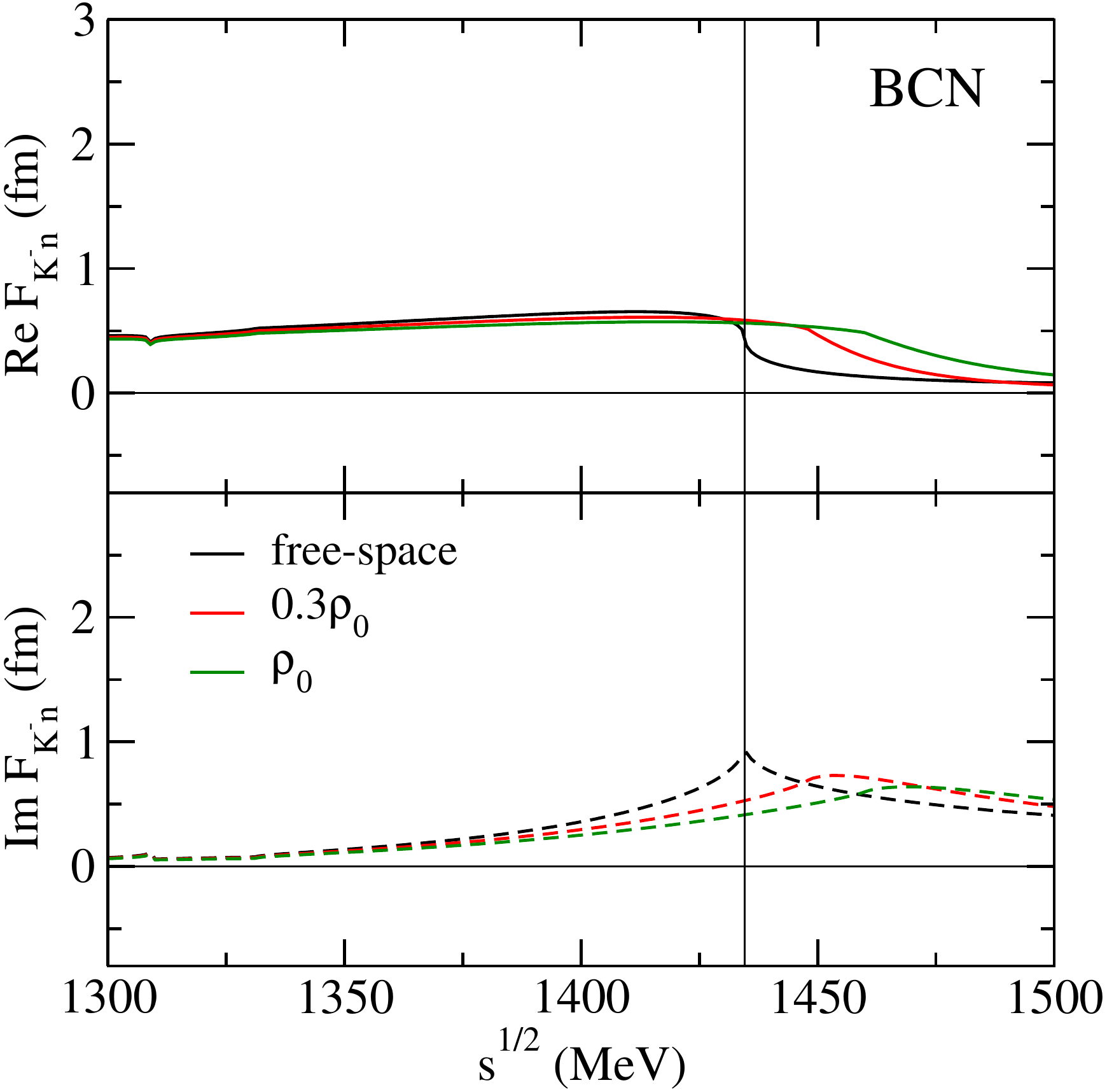} \hspace{10pt}
\includegraphics[width=0.48\textwidth]{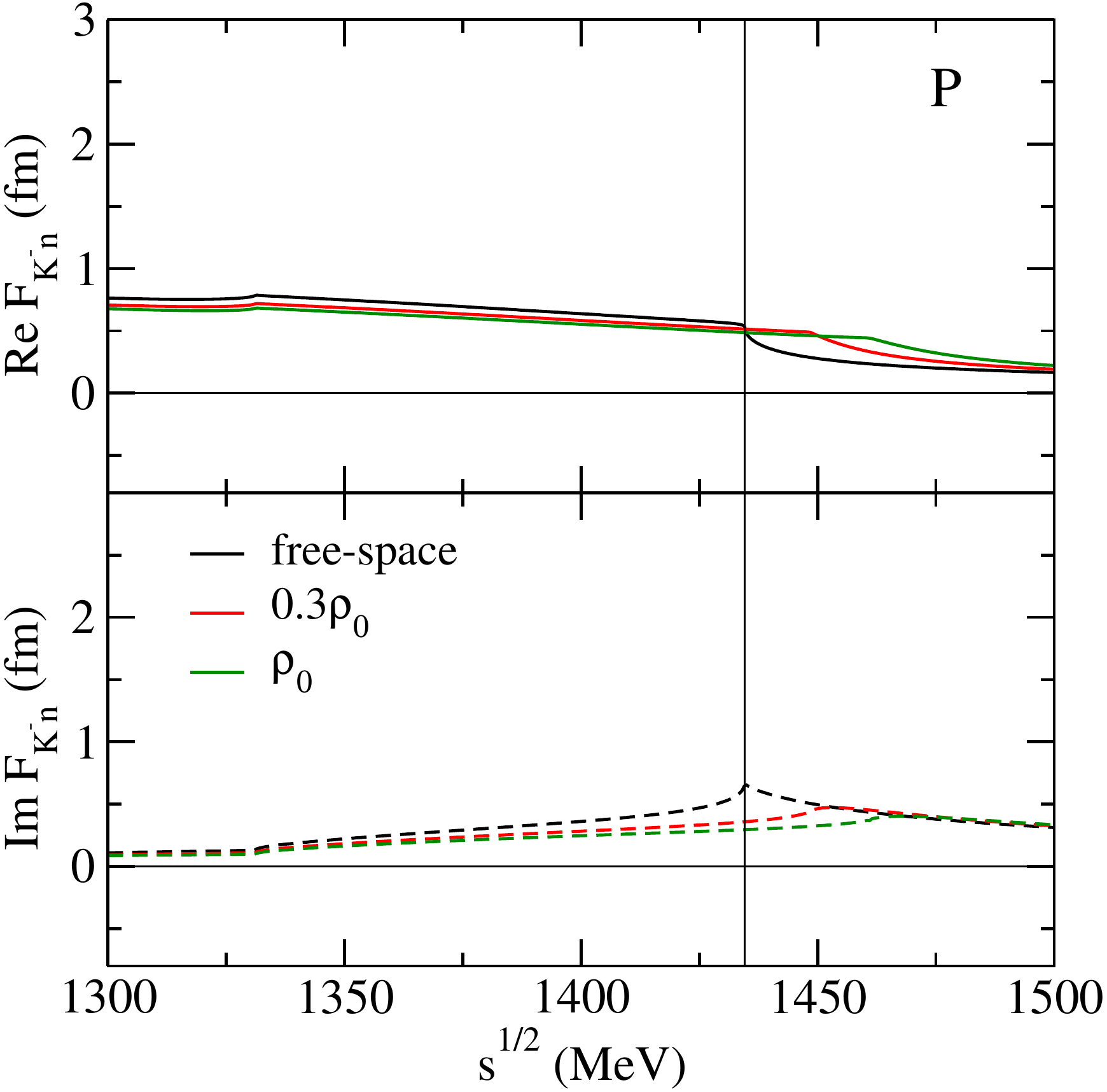}
\end{center}
\caption{The comparison of free-space (black) and Pauli blocked $K^-p$ (top) and $K^-n$ (bottom) amplitudes at two different densities: $0.3\rho_0$ (red) and $\rho_0$ (green), calculated in the BCN (left) and P (right) models.}
\label{fig:amplitudes}
\end{figure} 
   
The $K^-N$ t-matrices employed in our calculations are derived from two different chiral coupled channel meson-baryon interactions, namely the Barcelona model (BCN) \cite{bcn} and the Prague model (P) \cite{pnlo}. The parameters of both models are fitted to low-energy $K^-p$ data such as low-energy 
$K^-p$ scattering cross-sections \cite{kp_crosssection1, kp_crosssection2, kp_crosssection3}, threshold branching ratios \cite{kp_ratios1, kp_ratios2} and the strong interaction energy shift and width of kaonic hydrogen atom~\cite{SIDDHARTA}. The two models were also confronted with kaonic atom data and found to reproduce the $1N$ absorption fractions from bubble chamber experiments \cite{bubble1, bubble2, bubble3}, after adding a phenomenological $K^-$ multi-nucleon optical potential \cite{fgNPA17}.

Here we compare results for the free-space as well as Pauli blocked $K^-N$ amplitudes. The Pauli blocking effect is accounted for directly in the BCN and P models by restricting the nucleon momentum in the intermediate meson-nucleon loops of the unitarized amplitude to be larger than the Fermi momentum $k_F$. 

The free-space and Pauli blocked $K^-p$ (top panel) and $K^-n$ (bottom panel) amplitudes obtained from the BCN (left) and P (right) models are shown in Fig.~\ref{fig:amplitudes}, where we have defined:
\begin{equation}
    F_{K^-N}=-\frac{1}{4 \pi}\frac{m_N}{\sqrt{s}}~t_{K^-N\to K^-N}~,
\end{equation}
with $t_{K^-N\to K^-N}$ being the two-body t-matrix.

The $K^-p$ amplitudes are strongly energy dependent in both models due to the subthreshold resonance $\Lambda(1405)$. The BCN free-space $K^-p$ amplitude is smaller in magnitude than that of the P model at subthreshold energies. When the Pauli blocking effects are taken into account both models yield similarly reduced $K^-p$ amplitudes for the two considered densities. As the density increases, the resonant structures associated with the $\Lambda(1405)$ shifts above threshold. As a consequence, the free-space $K^-p$ interaction which is repulsive at threshold becomes attractive in the medium.  

The energy dependence of the mildly attractive $K^-n$ amplitudes is less pronounced for both models. The real part of the $K^-n$ amplitude in the BCN model tends to decrease with decreasing energies below threshold, contrary to the P model. The imaginary part of the $K^-n$ amplitude at threshold in the BCN model is larger than in the P model, however, further below threshold the amplitudes are very similar to each other in both models. The Pauli blocked amplitudes decrease again in magnitude with increasing density.

\begin{figure}[t!]
\begin{center}
\includegraphics[width=0.7\textwidth]{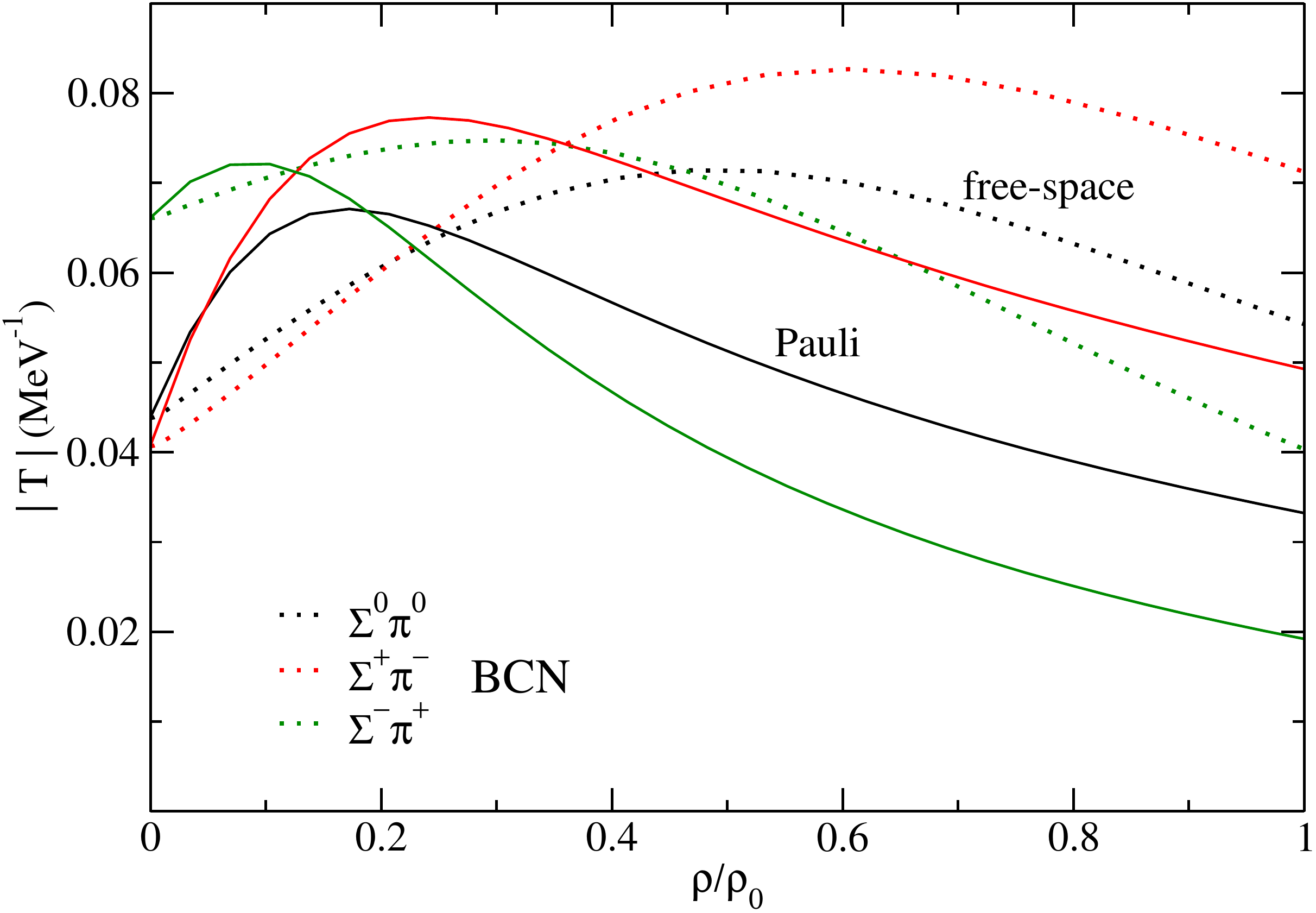}
\end{center}
\caption{Absolute values of $\Sigma^0 \pi^0$ (black), $ \Sigma^+ \pi^-$ (red), and $ \Sigma^- \pi^+$ (green) free-space (dotted) and Pauli blocked (solid) BCN model t-matrices as functions of relative density $\rho/\rho_0$, printed for energy shift corresponding to $B_{K^-}=0$~MeV and $p_{K^-}=0$~MeV/c.}
\label{fig:abs_t}
\end{figure}

In Fig.~\ref{fig:abs_t}, we compare the absolute values of the free-space (dotted) and Pauli blocked (solid) BCN t-matrices for channels $K^-p \rightarrow  \Sigma^0\pi^0, \Sigma^+\pi^- , \Sigma^-\pi^+$ as functions of relative density $\rho/\rho_0$, calculated for energy shift $\delta \sqrt{s}$ corresponding to $B_{K^-}=0$~MeV and $p_{K^-}=~0$~MeV/c. It is to be noted that the density dependence of the free-space amplitudes stems from the relation between the energy $\sqrt{s}$ and density in Eq.~\eqref{eq:sqrtS}. The peak in the free-space $I=0$ $\Sigma^0\pi^0 $ channel comes from the $\Lambda(1405)$ resonance and it is placed around $0.5\rho_0$, which corresponds to $\delta \sqrt{s} \sim -20$~MeV. The different position of the peak for the differently charged channels is due to the interference between $I=0$ and $I=1$ amplitudes, which is absent in $\Sigma^0\pi^0$, and of different sign for $\Sigma^+\pi^-$ and  $\Sigma^-\pi^+$. The medium modification of the amplitudes causes the peaks of the t-matrices to shift towards lower densities, $0.1 - 0.2 \rho_0$. Moreover, the absolute values of the Pauli blocked t-matrices decrease in magnitude with respect to the free-space ones for larger densities.

It is to be noted that self-energy insertions in terms of hadron-nucleon potentials for the intermediate baryons are not included in the in-medium amplitudes employed here. The self-energy effects are known to partly compensate for the Pauli upward shift of the amplitudes, moving them closer to the free-space ones \cite{Ramos:1999ku}. In this respect, the medium effects discussed in this work might be somehow modified. It will be worth employing in-medium amplitudes, which incorporate baryon as well as meson self-energy insertions, as soon as they become available for the BCN and P models. Being aware of this limitation, in the present work we will always refer to our in-medium effects as those associated to Pauli blocking correlations only.

\section{Results}
\label{results}

In this Section, we present the $K^-N$ and $K^-NN$ optical potentials and absorption branching ratios, calculated within the formalism described in Appendices~\ref{appendix:a} and \ref{appendix:b} and using the $K^-N$ scattering amplitudes derived from the BCN and P chiral models. 

First, we tested our model using the free-space amplitudes of Ref.~\cite{Oset_Ramos}, which were used in similar calculations in Ref.~\cite{sjPRC12}. We reproduce the result of Ref.~\cite{sjPRC12} for the $K^-N$ imaginary potential. In the case of the $K^-NN$ imaginary potential, we reproduce their result only if we adopt their prescription for the meson propagator of Eq.~(20), which is modified by the effect of short-range correlations and a monopole form-factor at each vertex. In the absence of these modifications, the resulting $K^-NN$ imaginary potential becomes twice as deep. Let us mention that the short-range correlations implemented in Ref.~\cite{sjPRC12} for all exchanged mesons were obtained following the phenomenological procedure of Ref.~\cite{Oset:1979bi} to deal with $NN$ correlations. However, here we are not facing a pure $NN$ interaction problem but a transition from a $NN$ pair to a $YN$ pair, the later probably having milder short-range correlations. Given this uncertainty, in the present work we have adopted a practical approach which consists in letting the form factor [Eq.~\eqref{eq:form_factor}] to account for these effects in an effective way. We have found that the results of the pion-exchange diagrams calculated with the short-range correlations of Ref.~\cite{sjPRC12}, Eq. (20), can be mimicked by our model employing a cut-off value $\Lambda_c=900$~MeV. The kaon-exchange diagrams, being of shorter-range due to the higher mass of the kaon, are more affected by correlations and require a somewhat softer cut-off value.
Therefore, our results will be presented in bands, obtained employing form factors with cut-off values in the range $\Lambda_c = 800-1200$~MeV, aiming in this way to account for the uncertainty tied to short-range baryon-baryon correlation effects.

Finally, we note that we do not consider the contribution from the $\Sigma(1385)$ resonance to $K^-$ absorption since for low-energy kaons it was found to be small in comparison with that coming from the $\Lambda(1405)$~\cite{sjPRC12}.

\subsection{$K^-N$ and $K^-NN$ optical potentials in nuclear matter}
\begin{figure}[t!]
\begin{center}
\includegraphics[width=0.48\textwidth]{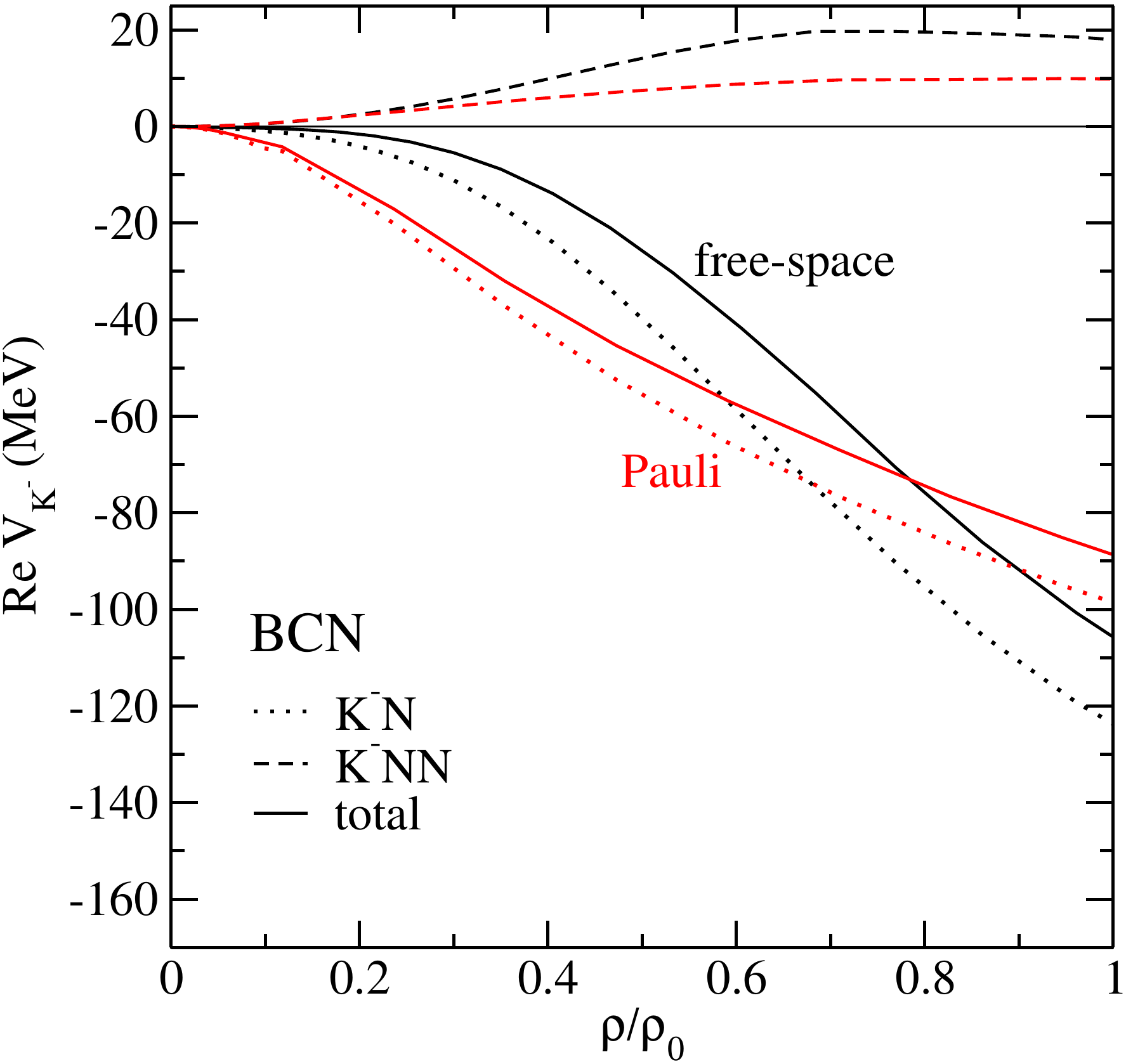} \hspace{10pt}
\includegraphics[width=0.48\textwidth]{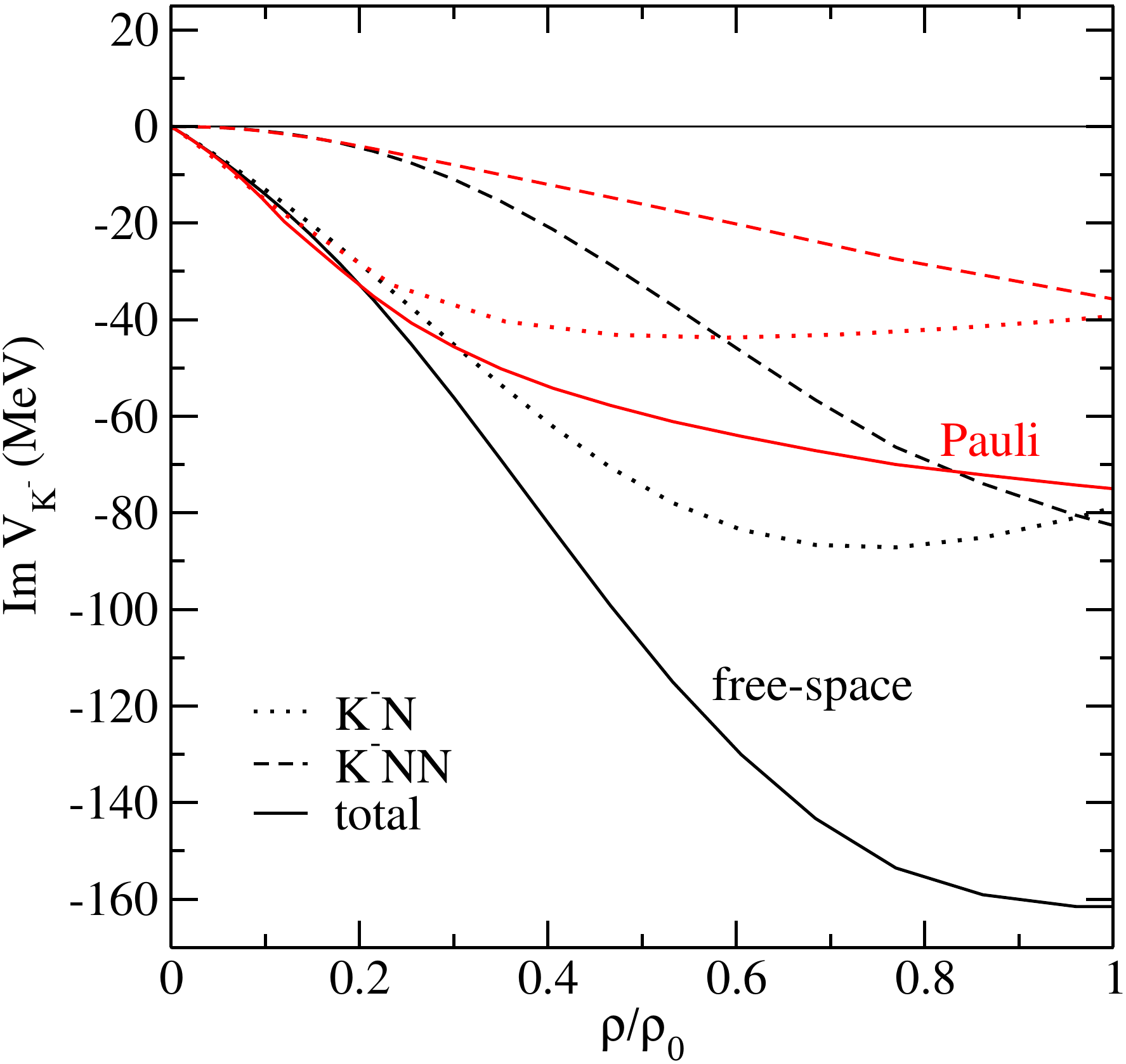}
\end{center}
\caption{The real (left) and imaginary (right) parts of the $K^-N$ (dotted), $K^-NN$ (dashed), and total (solid) optical potentials as a function of relative density $\rho/\rho_0$, calculated for $B_{K^-}\!=~\!0$~MeV and $p_{K^-}\!=\!0$~MeV/c using the free-space (black) and Pauli blocked (red) BCN amplitudes.}
\label{fig:tpot_fs_pauli}
\end{figure}
First, we present results calculated with a cut-off value $\Lambda_c=1200$~MeV. In Fig.~\ref{fig:tpot_fs_pauli} we demonstrate the importance of employing Pauli blocked amplitudes. The real (left) and imaginary (right) parts of the $K^-N$, $K^-NN$, and total $K^-$ optical potential ($K^-N+K^-NN$) are presented as functions of the relative density $\rho/\rho_0$, calculated for $B_{K^-}=0$~MeV and $p_{K^-}=0$~MeV/c using the free-space (black) and Pauli blocked (red) $K^-N$ amplitudes derived from the BCN model. The real part of the $K^-NN$ potential calculated with free-space amplitudes is repulsive in the whole density region. The depth of the corresponding imaginary $K^-NN$ potential increases with the density and reaches $\sim -80$~MeV at $\rho_0$. The $K^-NN$ absorption starts to be dominant over the $K^-N$ absorption for $\rho > 0.9\rho_0$. The total absorptive $K^-$ potential with free-space amplitudes is deeper than the corresponding real part in the entire density region. 

When the Pauli blocking effect is taken into account, the absorptive $K^-N$ and $K^-NN$ potentials are reduced by approximately one half at saturation density with respect to the free-space potentials. The minimum of the Pauli blocked $K^-N$ absorptive potential is reached at a lower density since, as can be seen in Fig.~\ref{fig:abs_t}, the resonant $\Lambda(1405)$ structure in the Pauli blocked $K^-p\to \Sigma\pi$ amplitudes has also moved to lower densities. The density dependence of the real $K^-$ single-nucleon potential changes its shape due to medium effects, which is also explained by the shifting of the Pauli blocked amplitudes. 
As for the real part of the $K^-NN$ potential, we find that the repulsive contribution is reduced by approximately one half. In general, a repulsive real $K^-NN$ potential is in line with most of the phenomenological $K^-$ multi-nucleon optical potentials consistent with kaonic atom data and $1N$ absorption fractions at threshold \cite{fgNPA17}. 
The total real and imaginary Pauli blocked $K^-$ potentials are of similar size around saturation density. Our results show that the Pauli correlations in the amplitudes have a pronounced effect on the $K^-$ absorption and should not be neglected. From now on we will present mainly the results calculated with the Pauli blocked amplitudes.

\begin{figure}[t!]
\begin{center}
\includegraphics[width=0.48\textwidth]{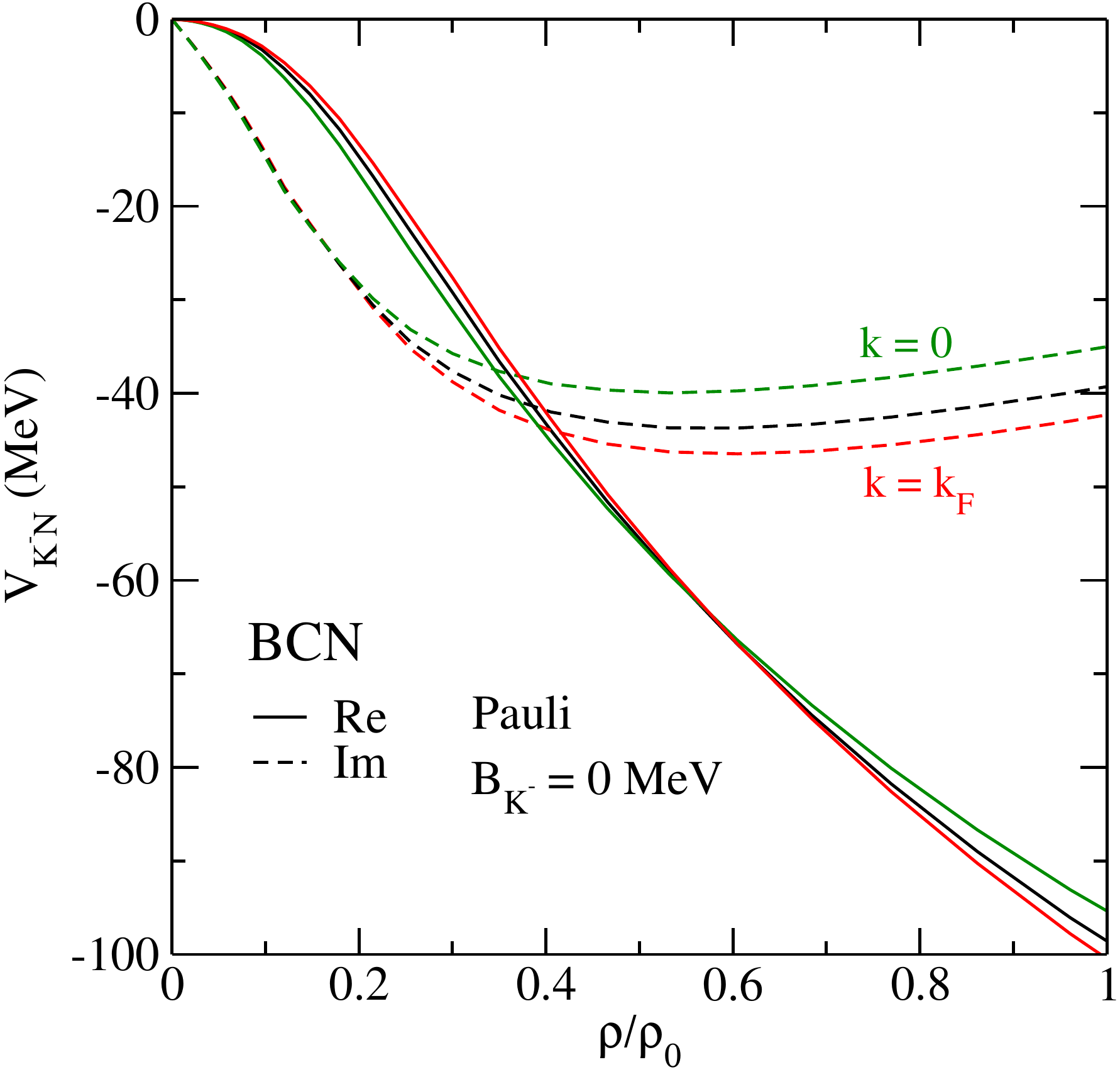} \hspace{10pt}
\includegraphics[width=0.47\textwidth]{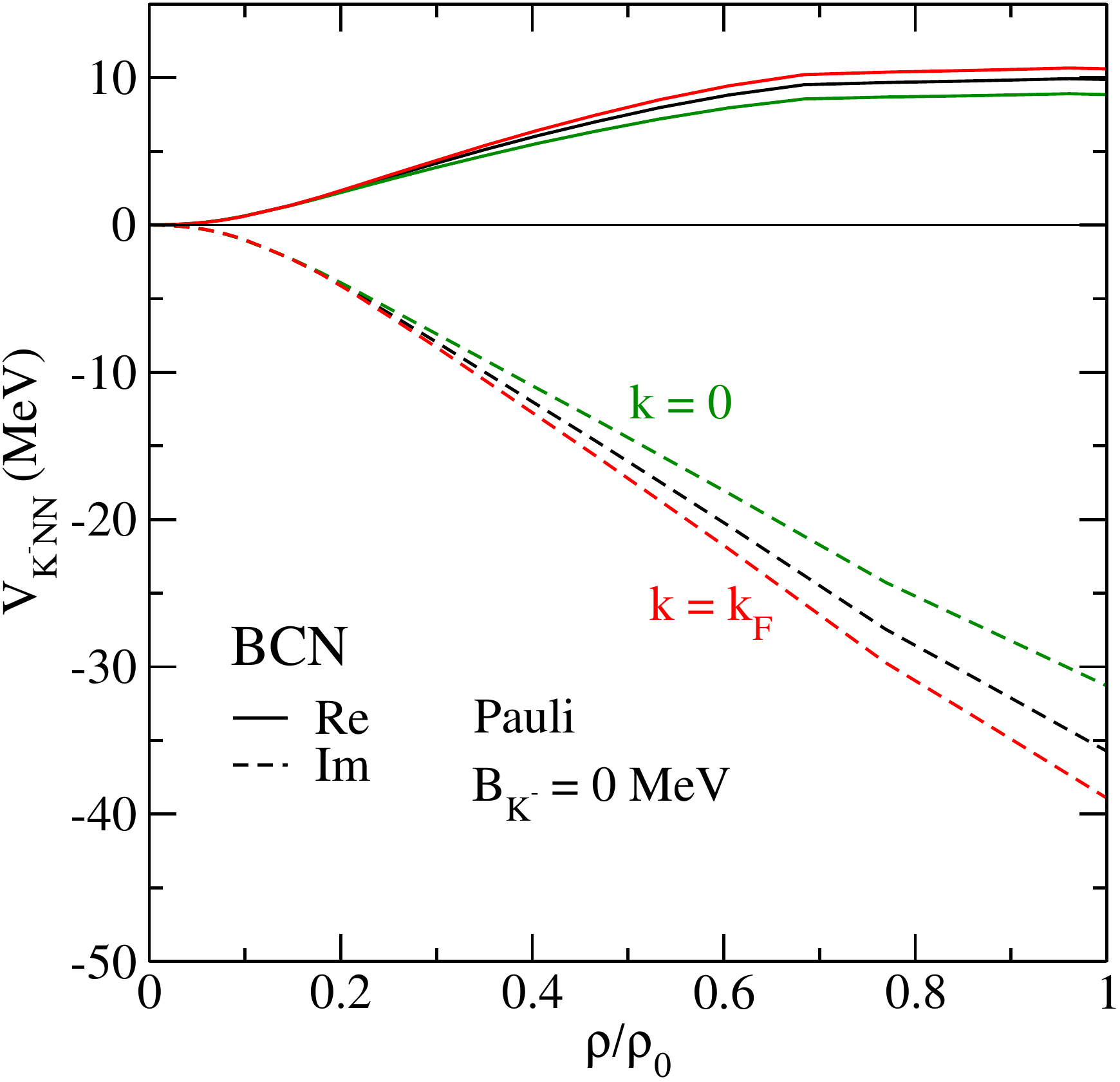}
\end{center}
\caption{The real (solid) and imaginary (dashed) parts of the $K^-N$ (left) and $K^-NN$ (right) optical potentials as a function of relative density $\rho/\rho_0$, calculated for $B_{K^-}=0$~MeV, $p_{K^-}=0$~MeV/c, and $\langle k \rangle=\sqrt{3/5}k_F$ (black) using the Pauli blocked BCN amplitudes compared with two extreme cases for the nucleon momentum $\vec{k} = 0$ (green) and $\vec{k} = k_F$ (red) in the expression for $\sqrt{s}$.}
\label{fig:fermi_motion}
\end{figure}

It is to be noted that in the expression for the $K^-N$ (see Appendix A) and $K^-NN$ (see Appendix B) potentials we approximate the nucleon momentum $\vec{k}$ by an average Fermi momentum, $\langle k \rangle=\sqrt{3/5} k_F$, which allows us to factorize the $K^-N$ t-matrices out of the integral. Since the $K^-p$ amplitudes are significantly energy dependent (see Fig.~\ref{fig:amplitudes}) and their argument $\sqrt{s}$ depends on the nucleon momentum one might argue that such approximation is not well justified. Therefore, we have calculated the single- and two-nucleon $K^-$ potentials considering two extreme cases for the nucleon momentum, $\vec{k} = 0$ and $\vec{k} = k_F$, in $\sqrt{s}$ [see Eq.~\eqref{eq:sqrtS}] to analyze the effect of averaging the nuclear Fermi motion. The results are shown in Fig.~\ref{fig:fermi_motion}. The $K^-$ potentials evaluated with an average nucleon momentum lie between the two extreme lines and the difference between potentials calculated for $\vec{k} = 0$ and $\vec{k} = k_F$ is up to $10$~MeV at $\rho_0$. This result suggests that a potential obtained by integrating over all possible values of nucleon momenta $\vec{k}$ in Eq.~\eqref{eq:sqrtS} would most likely be very close to the potential evaluated with an average nucleon momentum. 

\begin{figure}[t!]
\begin{center}
\includegraphics[width=0.48\textwidth]{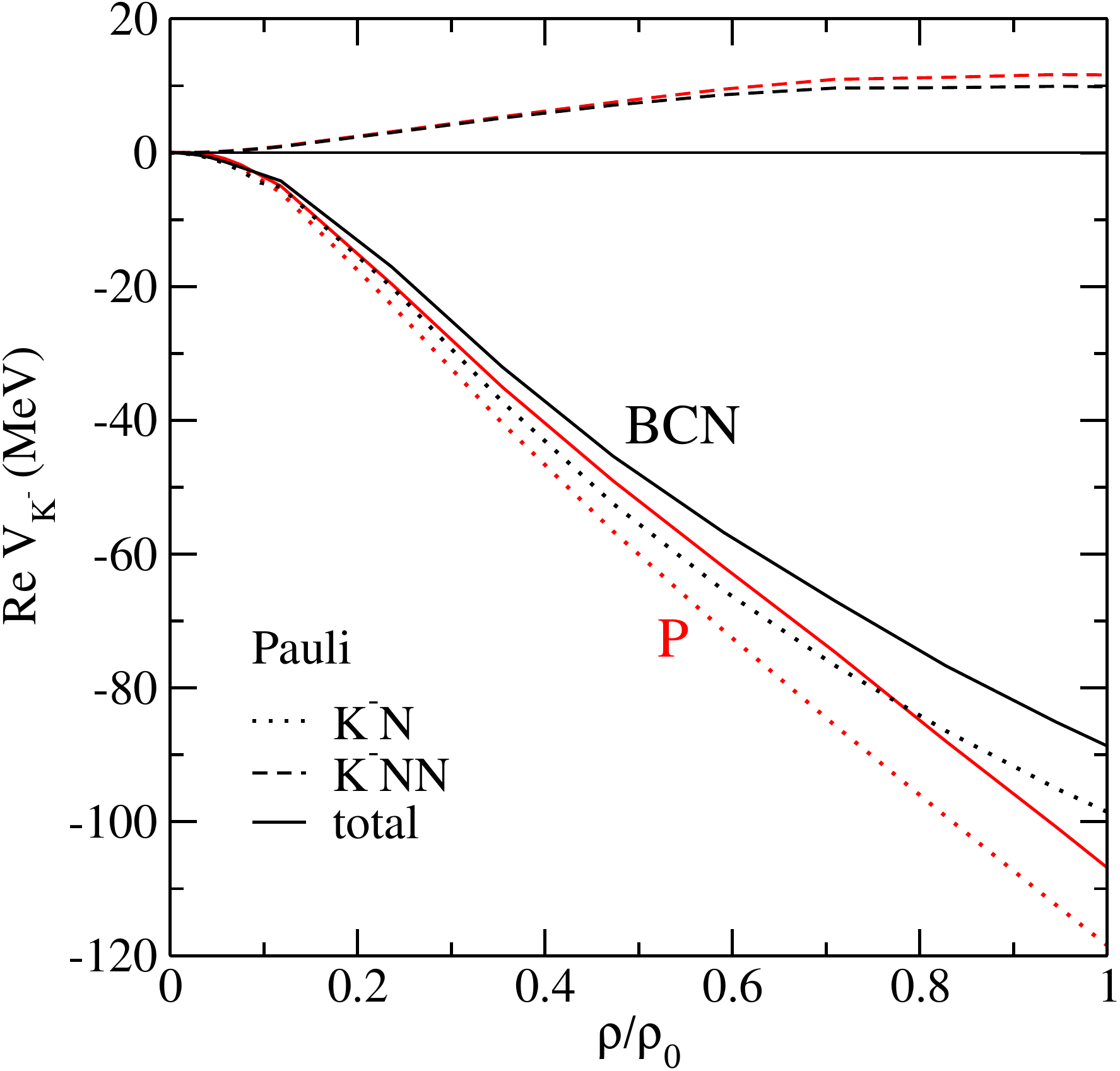} \hspace{10pt}
\includegraphics[width=0.48\textwidth]{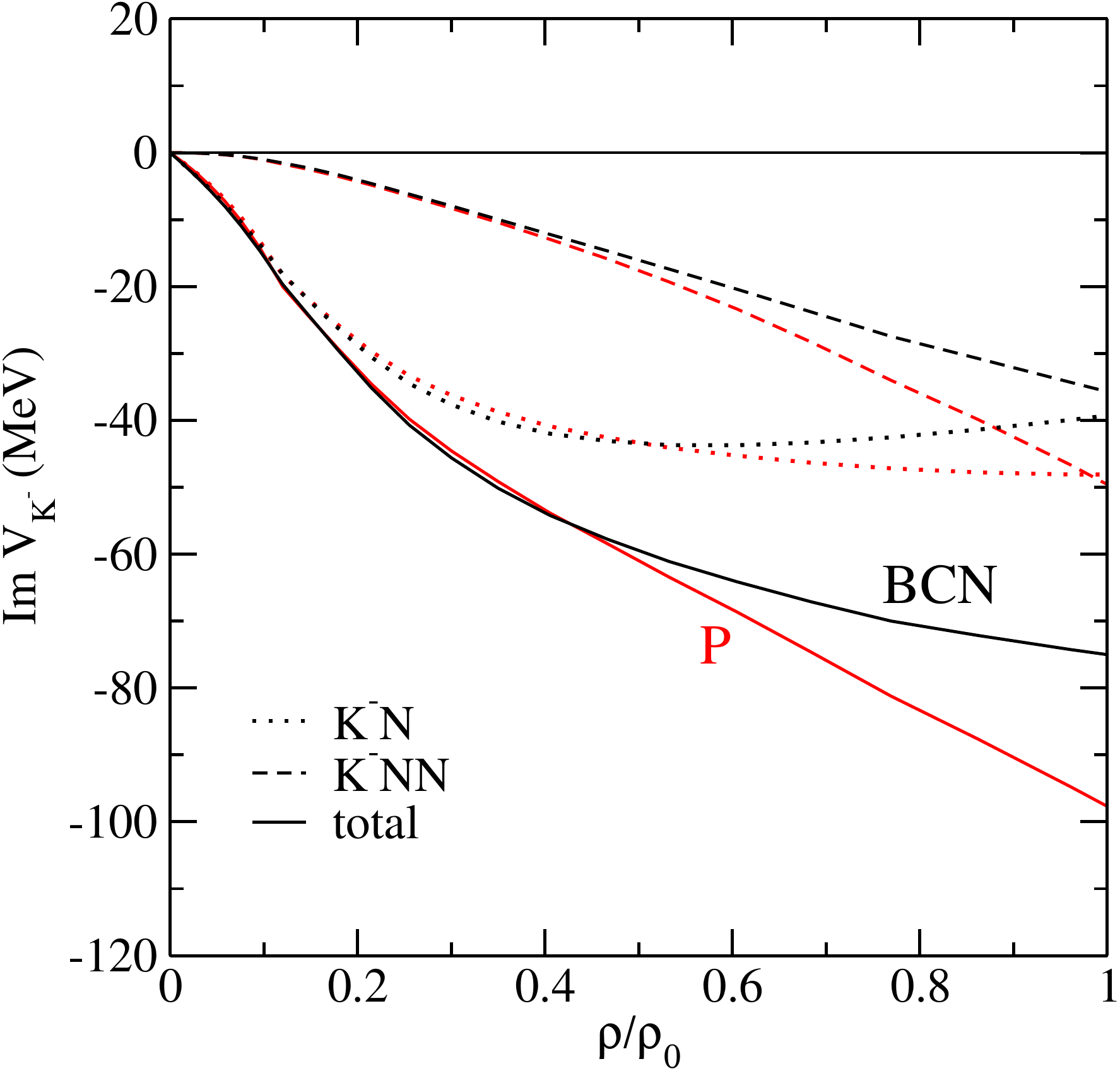}
\end{center}
\caption{Comparison of the real (left) and imaginary (right) parts of the $K^-N$ (dotted), $K^-NN$ (dashed), and total (solid) optical potentials as a function of relative density $\rho/\rho_0$, calculated for $B_{K^-}=0$~MeV and $p_{K^-}=0$~MeV/c using the Pauli blocked amplitudes within the BCN (black) and P (red) models.}
\label{fig:tpot_P_BCN}
\end{figure}

A comparison of the $K^-$ potentials obtained within the BCN and P models for $B_{K^-}=~0$~MeV and $p_{K^-}=0$~MeV/c is presented in Fig.~\ref{fig:tpot_P_BCN}. Both models yield qualitatively very similar $K^-$ potentials. The imaginary parts overlap up to $\sim 0.4\rho_0$. At $\rho_0$, the BCN model is less absorptive by about 20 MeV than the P model, this difference being equally shared by the $K^-N$ and $K^-NN$ contributions. As for the real part of the optical potential, we observe that the $K^-NN$ term is repulsive and of similar magnitude in both models. On the other hand, the P model yields a deeper real $K^-N$ potential than the BCN model, reaching a difference of about 20 MeV around saturation density.  

\begin{figure}[t!]
\begin{center}
\includegraphics[width=0.48\textwidth]{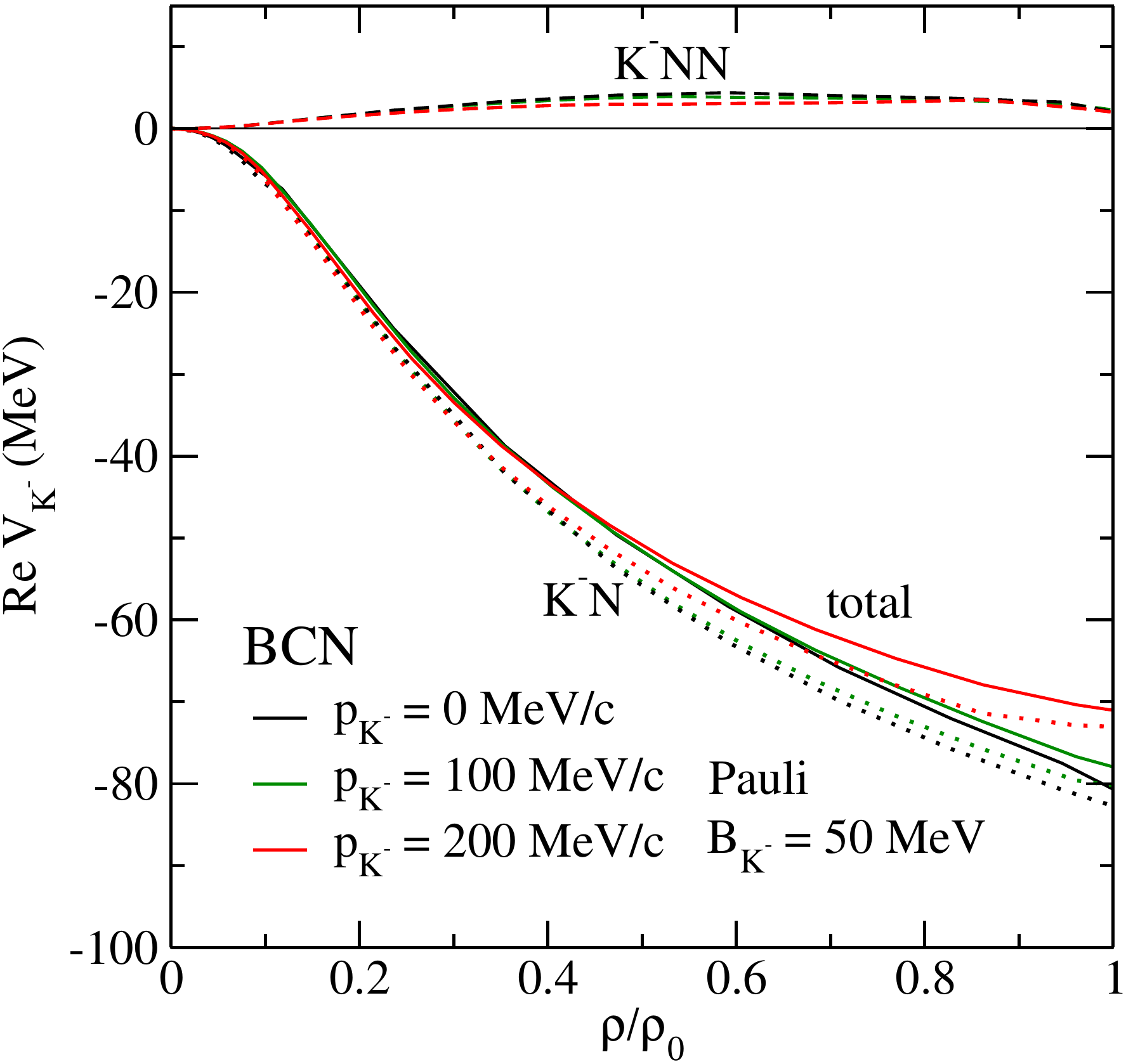} \hspace{10pt}
\includegraphics[width=0.48\textwidth]{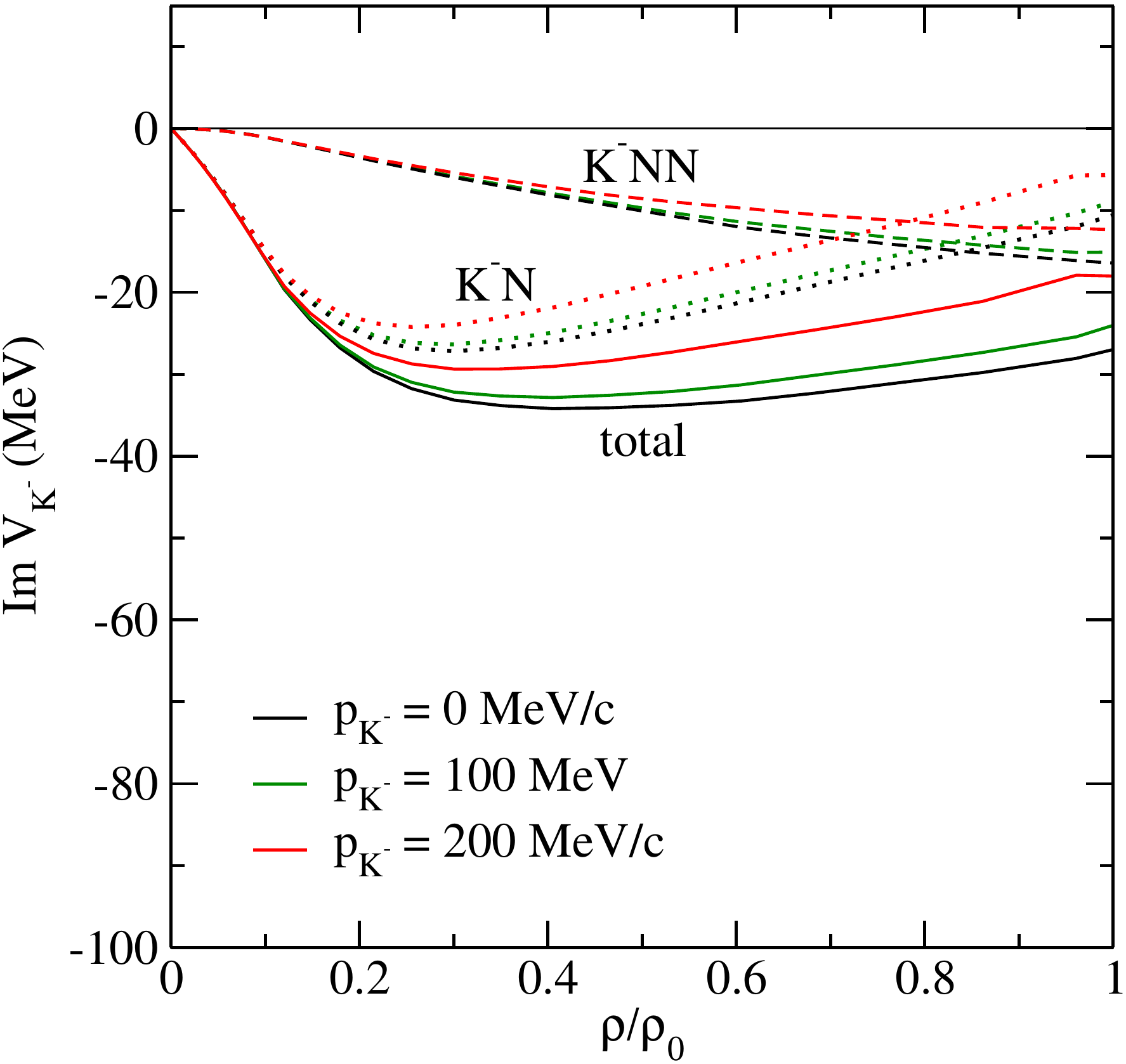}
\end{center}
\caption{The real (left) and imaginary (right) parts of the $K^-N$ (dotted), $K^-NN$ (dashed), and total (solid) optical potentials as a function of relative density $\rho/\rho_0$, calculated for $B_{K^-}=50$~MeV at $\rho_0$ and three different kaon momenta $p_{K^-}$ using the Pauli blocked BCN amplitudes.}
\label{fig:tpot_bk}
\end{figure}

Next, we consider the effect of a finite value of the $K^-$ binding energy $B_{K^-}$ and the kaon momentum $p_{K^-}$. In Fig.~\ref{fig:tpot_bk}, we present a comparison of the $K^-N$, $K^-NN$ and total $K^-$ optical potentials calculated for $B_{K^-}=50$~MeV at $\rho_0$, which probes center-of-mass $K^-N$ energies $\sqrt{s}$ down to 100~MeV below threshold (see Fig.~\ref{fig:deltaEs}), and three values of kaon momentum $p_{K^-}=0$~MeV/c (black), 100~MeV/c (green) and 200~MeV/c (red) at $\rho_0$ within the BCN model. First, we focus on the case of $p_{K^-}=0$~MeV/c (black). In the right panel, we observe that the imaginary $K^-N$ and $K^-NN$ potentials for the binding energy $B_{K^-}=50$~MeV are shallower than those obtained for $B_{K^-}=0$~MeV (see Fig.~\ref{fig:tpot_P_BCN}). This is because the $B_{K^-}=50$~MeV case explores the $K^-p\to \Sigma\pi$ amplitudes at energies deeper below the $K^- N$ threshold, i.~e. further away of the resonant peak and closer to the $\pi \Sigma$ threshold (see Fig.~\ref{fig:deltaEs}). The $K^-N$ imaginary potential decreases faster with the density then the $K^-NN$ absorptive potential. As a results, the $K^-NN$ imaginary potential becomes deeper than the $K^-N$ potential at $\rho > 0.8 \rho_0$. The depth of the total $K^-$ absorptive potential obtained for $B_{K^-}=50$~MeV is reduced by more than one half at $\rho_0$ with respect to the $B_{K^-}=0$~MeV case. The real part of the $K^-NN$ optical potential (left panel) becomes substantially reduced when a finite antikaon binding energy is employed. On the other hand, the total real $K^-$ potential, which is dominated by the $K^-N$ contribution, is little affected. When a finite value of the kaon momentum is taken into account the real and imaginary $K^-$ potentials decrease in magnitude as the value of the momentum increases. This happens because the non-zero value of $p_{K^-}$ causes a downward energy shift (see Fig.~\ref{fig:deltaEs}), thus probing a lower energy region where the $K^-N$ amplitudes are smaller (see Fig.~\ref{fig:amplitudes}). The effect is most pronounced around $\rho_0$. Similar trends in the behaviour of the $K^-$ potentials are observed also for $B_{K^-}=0$~MeV and finite values of kaon momentum. The potentials calculated with the P model (not shown in the figure) exhibit the same features.

\begin{figure}[t!]
\begin{center}
\includegraphics[width=0.48\textwidth]{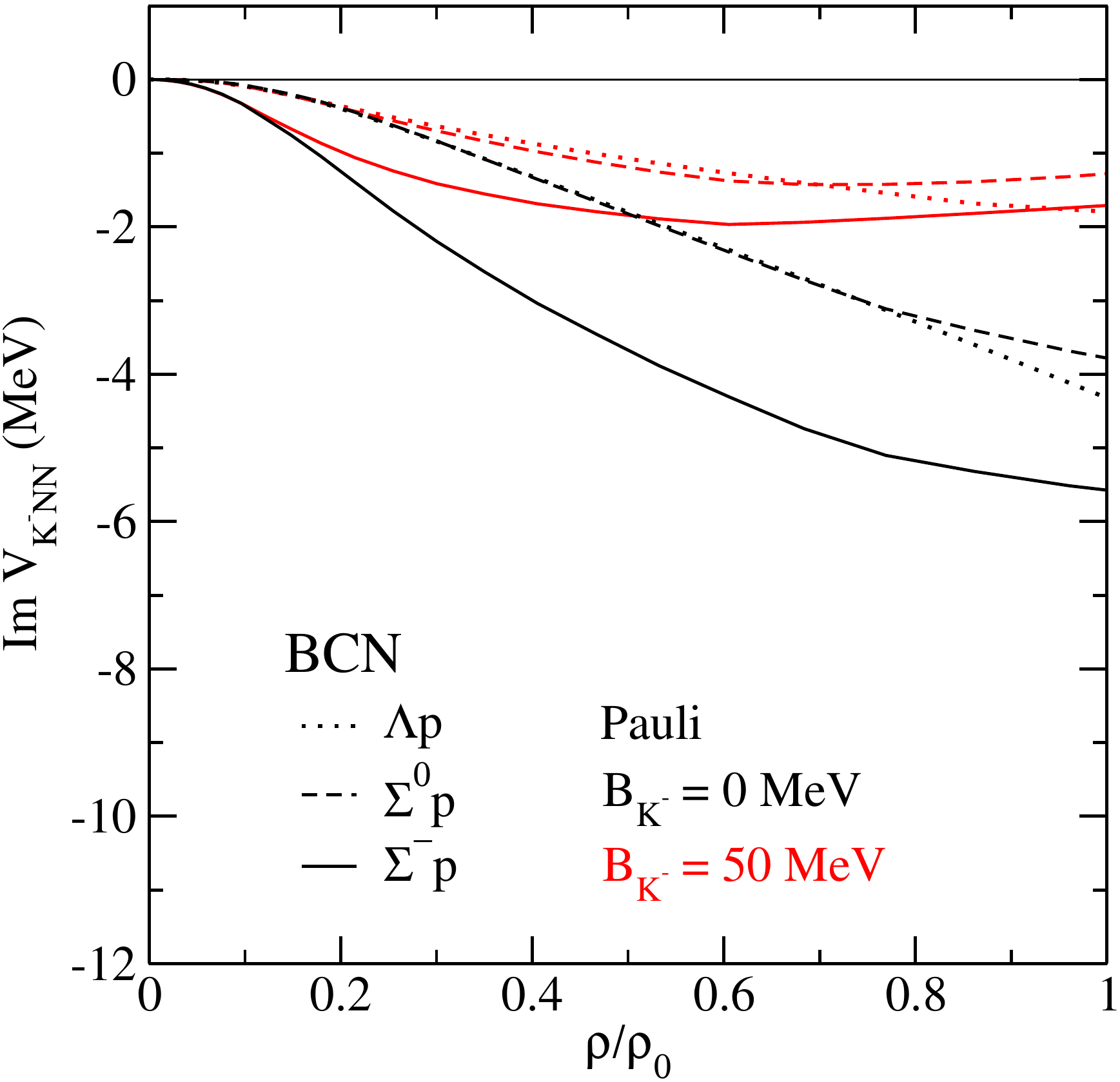} \hspace{10pt}
\includegraphics[width=0.48\textwidth]{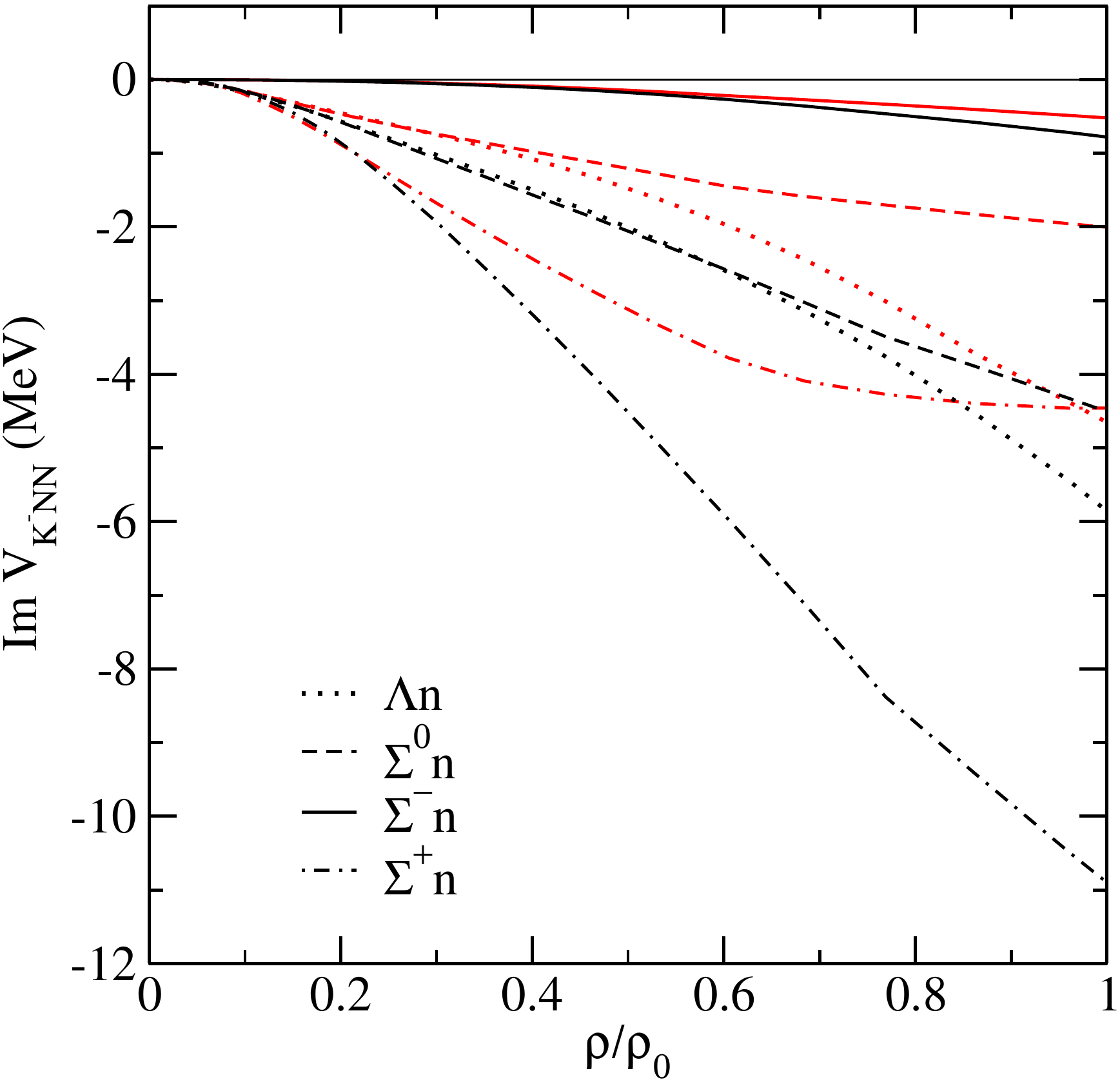}
\end{center}
\caption{Respective contributions to the $K^-NN$ absorptive potential corresponding to $\Lambda(\Sigma)p$ (left) and $\Lambda(\Sigma)n$ (right) final states, calculated for $p_{K^-}=~0$~MeV/c and $B_{K^-}=~0$~MeV (black) and $50$~MeV at $\rho_0$ (red) using the Pauli blocked BCN amplitudes.}
\label{fig:contributions_knn_bk}
\end{figure}

In Fig.~\ref{fig:contributions_knn_bk} we plot the $K^-$ two-nucleon annihilation contributions into different final states as functions of the relative density $\rho/\rho_0$, calculated for $p_{K^-}=0$~MeV/c in the case of $B_{K^-}=0$~MeV (black) and $50$~MeV (red) within the BCN model. The left panel shows the $K^-NN$ absorptive potential corresponding to the $Yp$ ($Y=\Lambda, \Sigma^0, \Sigma^-$) channels and the right panel that of the $Yn$ ($Y=\Lambda, \Sigma^0, \Sigma^-, \Sigma^+$) ones. The  $K^-NN$ absorptive contributions are, in general, significantly reduced when $B_{K^-}$ is increasing from 0 to 50 MeV, except for the $K^-nn\rightarrow \Sigma^-n$ case, which also gives the smallest contribution. This is due to the fact that this absorption process involves pure isospin $I=1$ $K^-n$ amplitudes only, much more moderate in size and less energy dependent than the $\Lambda(1405)$ dominated isospin $I=0$ ones which play an important role in the other channels.  

Now let us focus on the remaining contributions to the $\Sigma N$ final states in the case of $B_{K^-}=0$~MeV. We first notice that the $\Sigma N$ channels are dominated by pion-exchange, with the $\Sigma$ hyperon emitted from the kaon absorption vertex and the virtual pion absorbed in a Yukawa-type $NN\pi$ vertex. The reason lies in the fact that the kaon-exchange mechanism involves a Yukawa $\Sigma N K$ vertex which has a much reduced strength with respect to the pionic one, as noted already in \cite{sjPRC79}. This pion-exchange dominance explains why the contributions of the $\Sigma^+ n$ and $\Sigma^- p$ channels are much larger than the $\Sigma^0 p$ and $\Sigma^0 n$ ones, since a factor of $2=(\sqrt{2})^2$ already comes from the larger Yukawa coupling of charged pions versus that of neutral ones. The remaining difference is to be found in the relative strengths of the $K^-p\to \pi\Sigma$ amplitudes. As seen in Fig.~\ref{fig:abs_t}, the dominant amplitude at normal nuclear matter density is that to the $\Sigma^+\pi^-$ state, followed by $\Sigma^0\pi^0 $ and finally by $\Sigma^-\pi^+ $. This explains the different relative size of the dominant $\Sigma^+n$ and $\Sigma^-p$ channels in the $K^-NN$ absorption.

Our results are qualitatively very similar to those found
in Ref.~\cite{sjPRC12}, although there are some quantitative differences. Our absorptive widths to neutral hyperon final states ($\Lambda p$, $\Sigma^0 p$, $\Lambda n$ and $\Sigma^0 n$) are roughly 20\% smaller, the $\Sigma^+ n$ contribution is very similar and we find a smaller $\Sigma^- p$ strength by about a factor two. Overall, as seen in Fig.~\ref{fig:tpot_P_BCN}, our total $K^- NN$ absorptive potential is $-38$ MeV at normal nuclear matter density, about 20\% smaller in size than the value found in Ref.~\cite{sjPRC12}. These differences are mainly due to the fact that we are employing Pauli blocked amplitudes as well as different $K^- N$ interaction models.

\begin{figure}[t!]
\begin{center}
\includegraphics[width=0.48\textwidth]{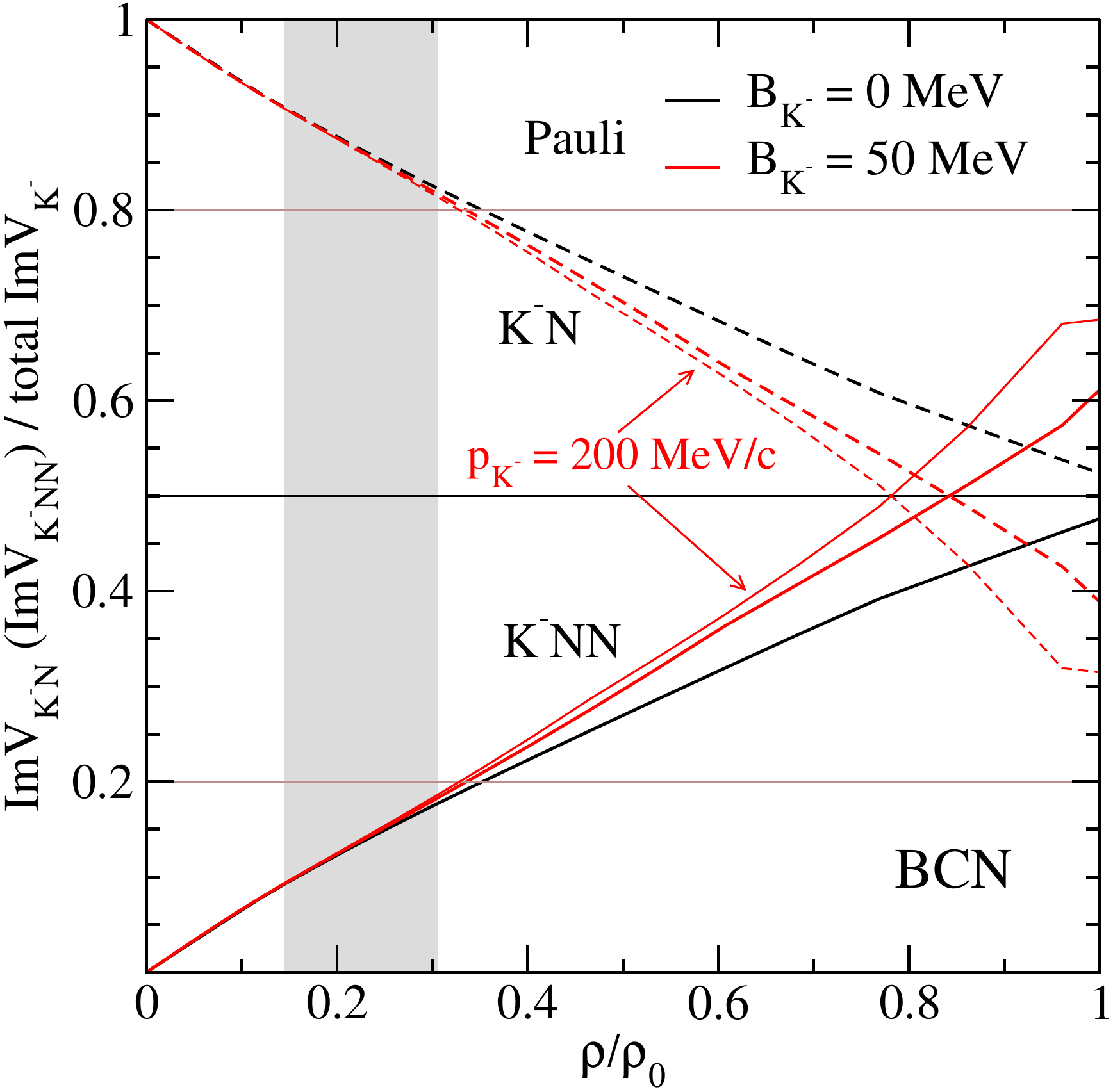} \hspace{10pt}
\includegraphics[width=0.48\textwidth]{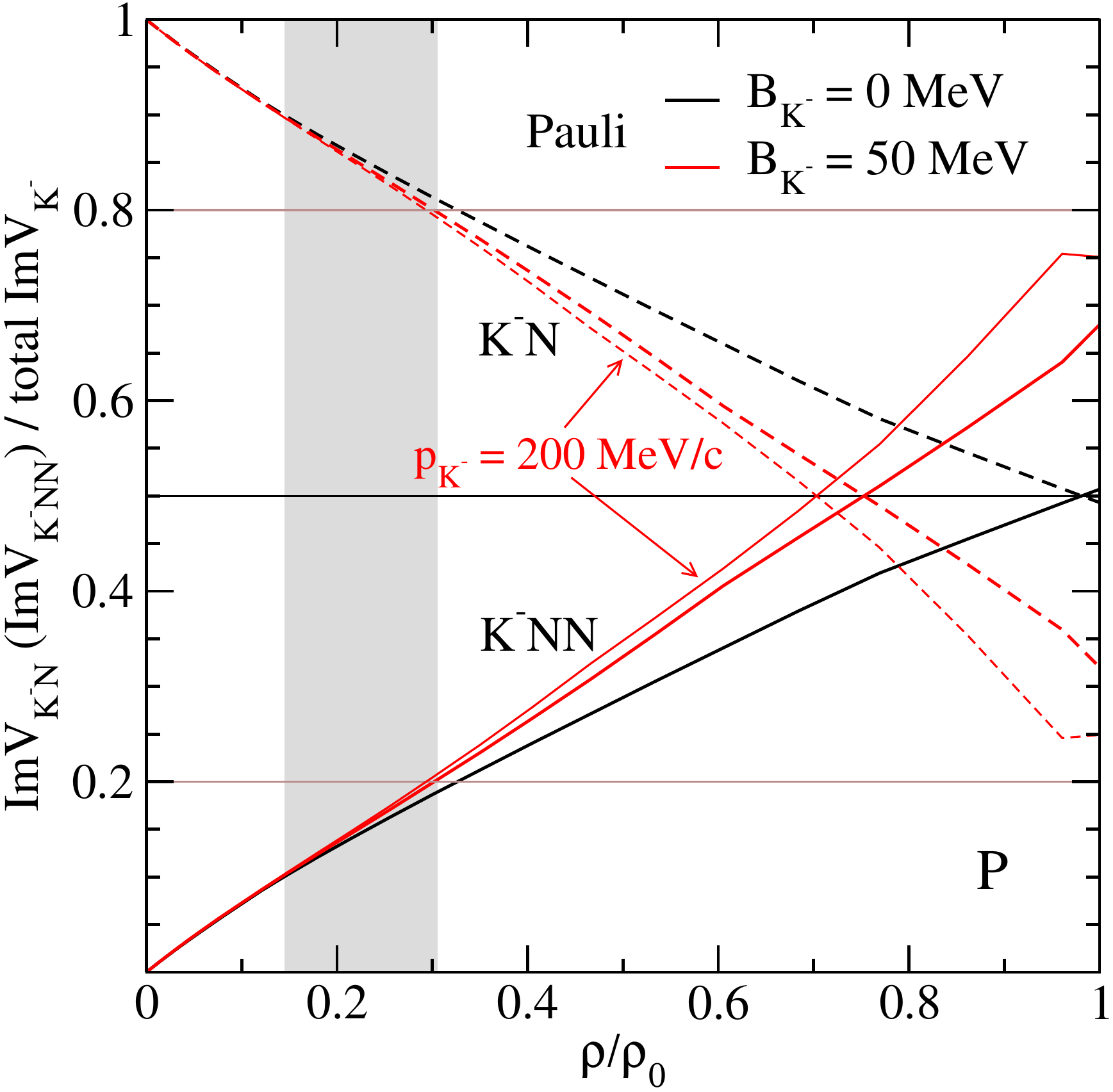}
\end{center}
\caption{Ratio of $K^-$ single-nucleon ($K^-N$) and two-nucleon ($K^-NN$) absorptive potentials to the total $K^-$ absorptive potential as a function of relative density $\rho/\rho_0$, calculated for $B_{K^-}\!=\!0$~MeV (black), $B_{K^-}\!=\!50$~MeV (thick red) with $p_{K^-}\!=\!0$~MeV/c, and for $B_{K^-}\!=\!50$~MeV and $p_{K^-}\!=\!200$~MeV/c at $\rho_0$ (thin red) using the BCN (left) and P (right) Pauli blocked amplitudes. The gray band denotes the region of densities probed in experiments with low-energy~$K^-$.}
\label{fig:ratio_kn_knn_PB}
\end{figure}

The ratios of $K^-$ single-nucleon and two-nucleon absorption widths to the total $K^-$ width in nuclear matter are shown in Fig.~\ref{fig:ratio_kn_knn_PB} as functions of the relative density $\rho/\rho_0$, calculated for $p_{K^-}=0$~MeV/c, in the case of $B_{K^-}=0$~MeV (black) and $50$~MeV (red) within the BCN (left panel) and P (right panel) models. The ratios for $p_{K^-}=200$~MeV/c and $B_{K^-}=50$~MeV at $\rho_0$ (thin red lines) are shown for comparison. The relative strength of the $K^-$ single-nucleon absorption decreases with density, while that of the two-nucleon term increases. Finally, the absorption of $K^-$ on two nucleons prevails. This is due to the reduction of phase space for the $K^-N\rightarrow \pi Y$ absorption channels in the vicinity of the $\pi\Sigma$ threshold.   
We find similar results for both chiral models. For $B_{K^-}=0$~MeV, the two ratios cross each other at or slightly above $\rho_0$, differently to what is found in Ref.~\cite{sjPRC12} where these ratios do not cross at all, not even up to the density of $0.2$~fm$^{-3}$ explored there. The reason is found again in that we employ Pauli blocked amplitudes. As seen in Fig.~\ref{fig:abs_t},  the $K^-p\to \Sigma\pi$ Pauli blocked amplitudes show the $\Lambda(1405)$ resonant structure at lower densities, while at higher densities their magnitude is substantially smaller than the free-space ones, thereby producing a reduced one-nucleon absorption ratio. The crossing occurs at lower density when a finite value of the $K^-$ binding energy $B_{K^-}=50$~MeV is considered, and is shifted even more for $p_{K^-}=200$~MeV/c. 
For both models, the $K^-$ single-nucleon and two-nucleon absorption fractions at $0.3\rho_0$ are close to 80\% and 20\%, respectively, and there is a tiny difference between the ratios calculated for $B_{K^-}=0$~MeV and $50$~MeV at $\rho_0$ or for $p_{K^-}=0$~MeV/c and $p_{K^-}=200$~MeV/c at $\rho_0$ in the region of experimentally probed densities (gray band). Such behavior is consistent with bubble chamber data~\cite{bubble1, bubble2, bubble3} as well as with findings obtained in the $K^-$-bound state calculations with the phenomenological $K^-$ multi-nucleon potential~\cite{hmPLB_PRC}. Conversely, in Ref.~\cite{sjPRC12} the 1$N$- and $2N$-absorption ratios at $\rho\sim0.3\rho_0$ are around 90\% and 10\%, respectively.

\begin{figure}[t!]
\begin{center}
\includegraphics[width=0.48\textwidth]{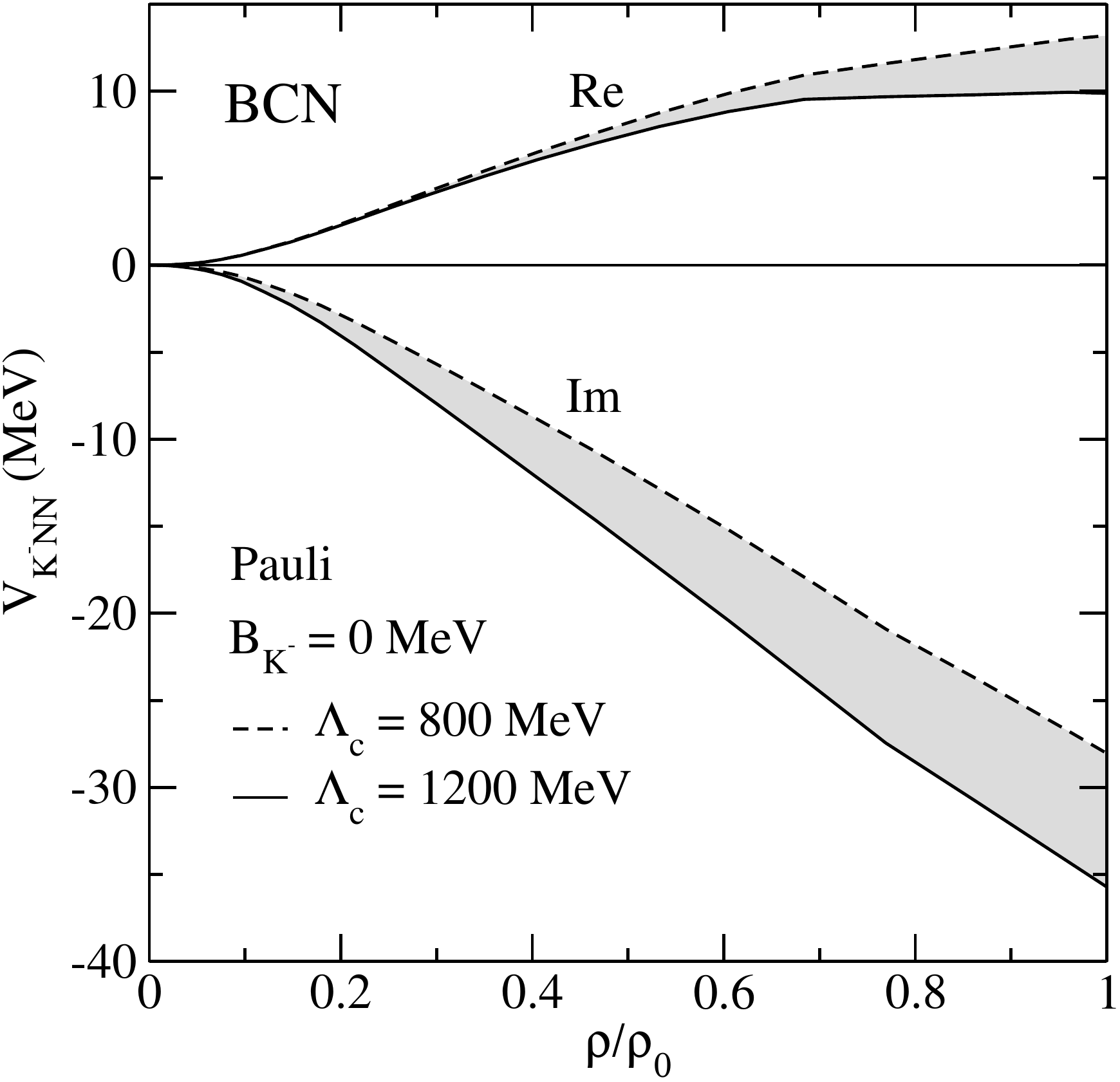}
\end{center}
\caption{Dependence of the $K^-NN$ potential on the value of the cut-off parameter $\Lambda_c$ used in the form factors, calculated for $B_{K^-}\!=\!0$~MeV and $p_{K^-}\!=\!0$~MeV/c with Pauli blocked BCN amplitudes.}
\label{fig:cut-off_dep}
\end{figure}

Finally, in Fig.~\ref{fig:cut-off_dep} we present the uncertainty (gray band) in the $K^-NN$ potential due to the range of cut-off values used in the form factor of Eq.~\eqref{eq:form_factor}, $\Lambda_c=800$ - $1200$~MeV, calculated for $B_{K^-}=0$~MeV and $p_{K^-}=0$~MeV/c using the Pauli blocked BCN amplitudes. When the lower cut-off value is used the depth of the imaginary potential decreases whereas the magnitude of the repulsive real potential increases.

\subsection{Comparison with experimental values of mesonic and non-mesonic absorption ratios}

Next, we present various ratios of single-nucleon (mesonic) and two-nucleon (non-mesonic) absorption widths, calculated within the BCN and P models. The ratios are calculated at $0.3\rho_0$ and $0.5\rho_0$ and compared with available experimental data. We consider $0.3\rho_0$ as the density region relevant for absorption of low-energy $K^-$ on $^{12}$C and $0.5\rho_0$ as a limiting density probed in experiments with low-energy $K^-$. Moreover, since the $K^-NN$ potential is dependent on the cut-off value $\Lambda_c$, we evaluate the branching ratios as averages of the ratios calculated for $\Lambda_c=800$~MeV and $\Lambda_c=1200$~MeV. The corresponding errors permit to reach the boundary values obtained for the two cut-offs.

\begin{table}[t!]
\caption\protect{Primary\footnote{Secondary interactions of the primary particles created in the absorption process were not considered.}-interaction branching ratios (in \%) for the $K^-$ two-nucleon absorption in nuclear matter, calculated with the BCN free-space and Pauli blocked amplitudes for $B_{K^-}=~0$~MeV and $p_{K^-}=0$~MeV/c. The errors denote the uncertainty due to the cut-off dependence.}
  \begin{tabular}{l|c|c||c|c}\label{tab:2N_channels}
   & \multicolumn{2}{c||}{$0.3 \rho_0$} & \multicolumn{2}{c}{$0.5 \rho_0$} \\ \hline
 BCN & ~free-space~ & ~Pauli~ & ~free-space~ & ~Pauli~ \\ \hline 
  $\Lambda n / K^-$  & 2.7 $\pm$ 0.5 & 1.9 $\pm$ 0.3 & 3.6 $\pm$ 0.5 & 2.8 $\pm$ 0.4  \\
  $\Lambda p /K^-$   & 1.4 $\pm$ 0.2 & 1.6 $\pm$ 0.3 & 2.6 $\pm$ 0.3 & 2.5 $\pm$ 0.3 \\
  $\Sigma^0 n / K^-$ & 2.6 $\pm$ 0.3 & 2.1 $\pm$ 0.3 & 3.7 $\pm$ 0.4 & 3.0 $\pm$ 0.3  \\
  $\Sigma^0 p / K^-$ & 1.3 $\pm$ 0.2 & 1.6 $\pm$ 0.2 & 2.5 $\pm$ 0.3 & 2.6 $\pm$ 0.3  \\
  $\Sigma^- n / K^-$ & 0.12 $\pm$ 0.01 & 0.12 $\pm$ 0.01 & 0.19 $\pm$ 0.01 & 0.25 $\pm$ 0.02 \\
  $\Sigma^- p / K^-$ & 5.7 $\pm$ 0.8 & 4.2 $\pm$ 0.6 & 7.9 $\pm$ 0.9 & 5.4 $\pm$ 0.6 \\
  $\Sigma^+ n / K^-$ & 2.8 $\pm$ 0.4 & 3.8 $\pm$ 0.5 & 5.6 $\pm$ 0.6 & 6.4 $\pm$ 0.7 
\end{tabular}
 
\end{table}

In Table~\ref{tab:2N_channels} we show the $2N$-absorption branching ratios into different final states. Similarly to the results shown in Fig.~\ref{fig:contributions_knn_bk}, the $\Sigma^+ n$ and $\Sigma^- p$ are the dominant channels, followed by the neutral hyperon ones ($\Lambda n,~\Lambda p,~\Sigma^0 n$ and $\Sigma^0 p$).
The presented values can by easily compared with experimental data as recent and future experiments aim at providing separate measurements of these ratios. 

\begin{figure}[tb!]
\begin{center}
\includegraphics[width=0.48\textwidth]{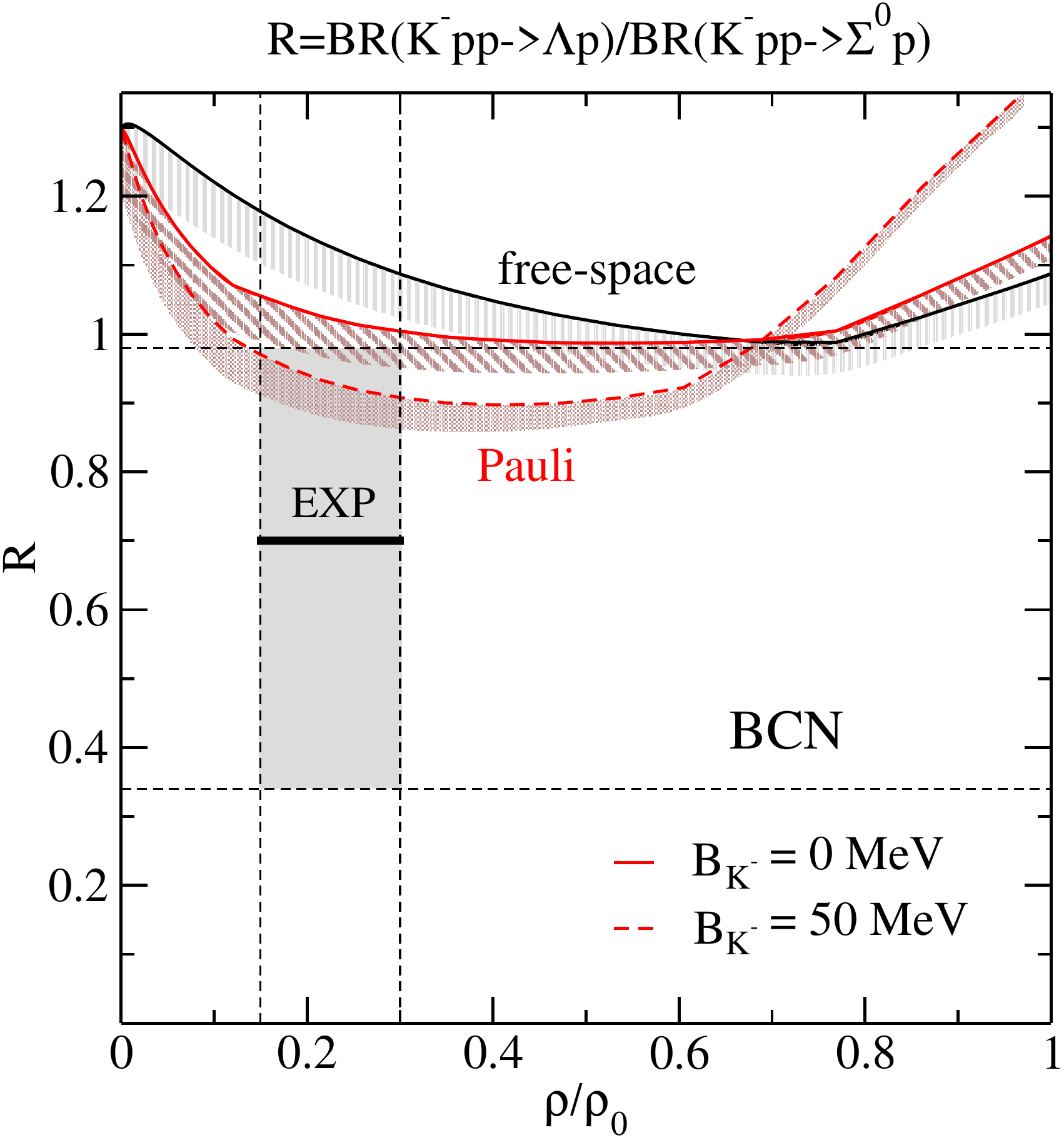} \hspace{10pt}
\includegraphics[width=0.48\textwidth]{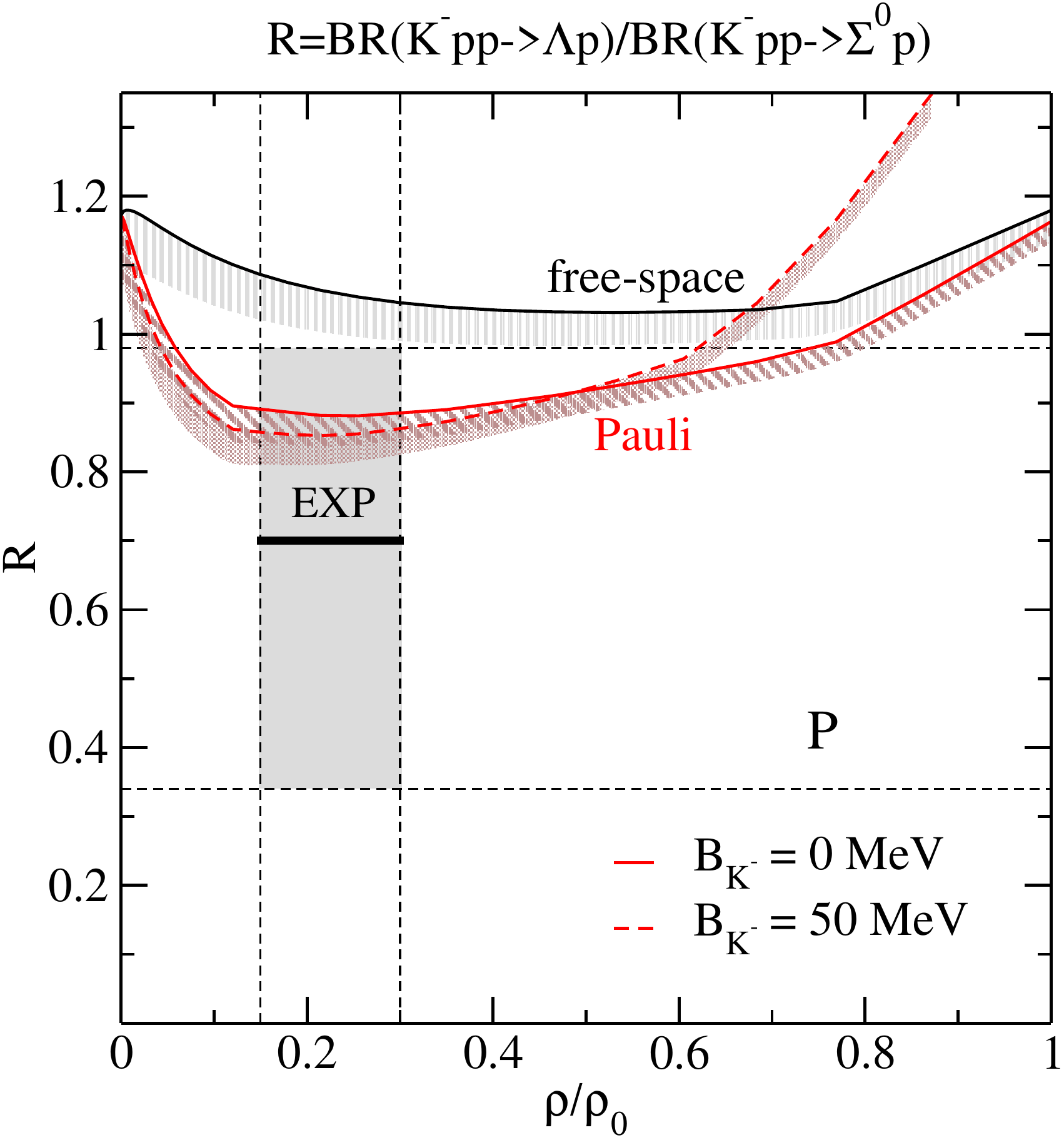}
\end{center}
\caption{The ratio of branching ratios for the $K^-pp\rightarrow \Lambda p$ and $K^-pp\rightarrow \Sigma^0 p$ channels as a function of relative density $\rho/\rho_0$, calculated for $p_{K^-}=0$~MeV/c and $B_{K^-}=0$~MeV and $50$~MeV at $\rho_0$ using the free-space (black) and Pauli blocked (red) BCN (left) and P (right) model amplitudes. Color bands denote the uncertainty due to different cut-off values $\Lambda_c=~800-1200$~MeV. The dashed vertical lines denote the region of densities probed in experiments with low-energy $K^-$ including the experimental value of the ratio with corresponding error bar (gray rectangle).}
\label{fig:ratio_Lp_S0p_PB}
\end{figure}

Let us now discuss the results for kaon absorption obtained in the recent counter-experiment of the AMADEUS collaboration that
measured $K^-$ multi-nucleon absorption fractions impinging low-energy $K^-$ produced at the DA$\Phi$NE collider on a carbon target~\cite{amadeus16,amadeus19}. They obtained the ratio of branching ratios
\begin{equation}\label{eq:LpSp_ratio}
R=\frac{{\rm BR}(K^-pp\rightarrow \Lambda p)}{{\rm BR}(K^-pp\rightarrow \Sigma^0 p)}=0.7 \pm 0.2({\rm stat.})^{+0.2}_{-0.3}({\rm syst.})~,
\end{equation}
for `quasi-free' production of $\Lambda(\Sigma^0)p$ pairs. These processes correspond to the direct emission of $\Lambda (\Sigma^0) p$ pairs and hence we can calculate their strength directly within our formalism. The corresponding ratios calculated using the BCN (left) and P (right) free-space (black lines) and Pauli blocked (red lines) amplitudes are plotted in Fig.~\ref{fig:ratio_Lp_S0p_PB} as functions of relative density $\rho/\rho_0$. The AMADEUS experimental value including error bar is shown for comparison. The lines denote the ratios calculated for the cut-off value $\Lambda_c=1200$~MeV and the bands denote the uncertainty due to lower values of $\Lambda_c$. The free-space ratio is similar to that obtained in Ref.~\cite{sjPRC12}. The Pauli blocking has a pronounced effect on the ratio, decreasing its value close to or below 1 for both chiral models, bringing it within the error bar of the experimental ratio at $0.15 - 0.3\rho_0$. Considering the $K^-$ binding energy $B_{K^-}=50$ MeV at $\rho_0$ further decreases the ratio $R$ in the region of experimentally accessible densities. Finally, we have checked that the non-zero value of kaon momentum has a negligible effect on the ratio $R$ in the region of experimentally relevant densities.

\begin{figure}[b!]
\begin{center}
\includegraphics[width=0.48\textwidth]{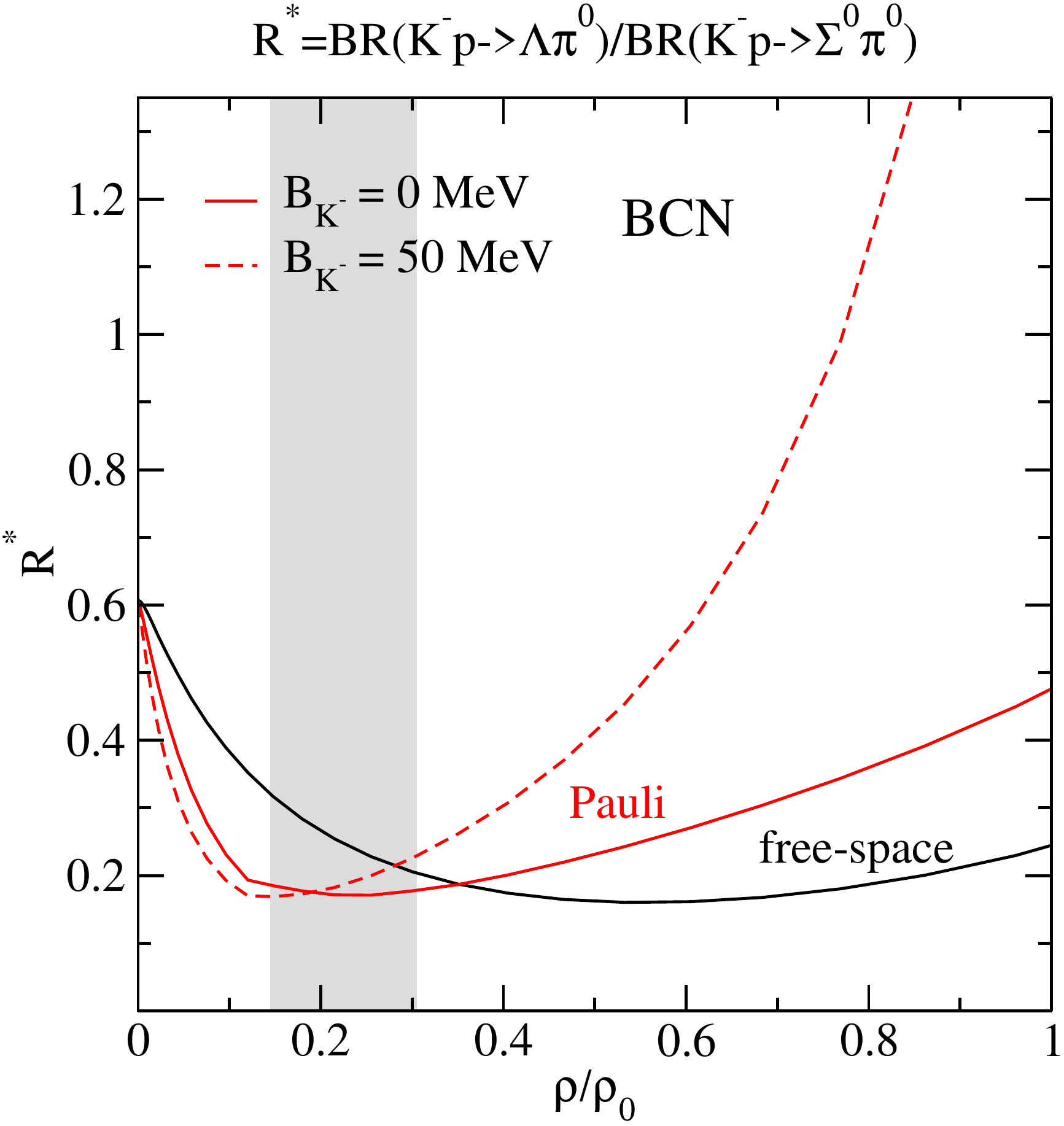} \hspace{10pt}
\includegraphics[width=0.48\textwidth]{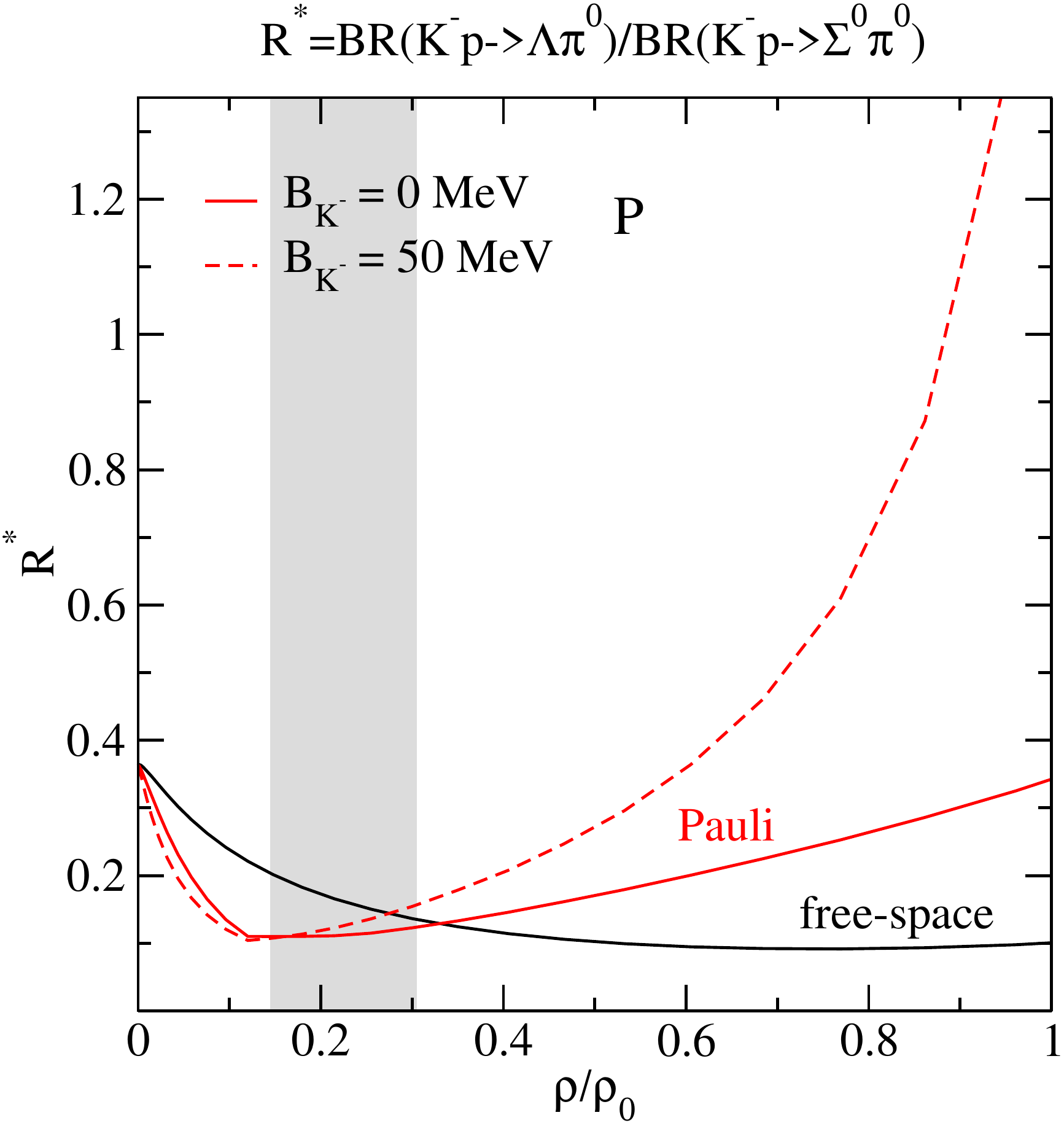}
\end{center}
\caption{The ratio of branching ratios for the $K^-p\rightarrow \Lambda \pi^0$ and $K^-p\rightarrow \Sigma^0 \pi^0$ channels as a function of relative density $\rho/\rho_0$, calculated for $p_{K^-}=0$~MeV/c and $B_{K^-}=0$~MeV and $50$~MeV at $\rho_0$ using the free-space (black) and Pauli blocked (red) BCN (left) and P (right) model amplitudes. The gray band denotes the region of densities probed in experiments with low-energy $K^-$.}
\label{fig:ratio_LpiSpi_PB}
\end{figure}

In the same experiment of Ref.~\cite{amadeus19}, it was argued that the ratio of Eq.~\eqref{eq:LpSp_ratio} for the $K^-pp\to\Lambda p,~\Sigma^0 p$ processes would be similar to the ratio for $K^-p$ absorption into $\Lambda\pi^0$ and $\Sigma^0\pi^0$ final states, given by
\begin{equation}\label{eq:LpiSpi_ratio}
R^*=\frac{{\rm BR}(K^-p\rightarrow \Lambda \pi^0)}{{\rm BR}(K^-p\rightarrow \Sigma^0 \pi^0)}~.
\end{equation}
The argument for the similarity of the two ratios relies in assuming that the most important contribution for both $2N$-absorption processes comes from $\pi^0$ exchange, on the basis of the dominance of the $\Lambda(1405)$ resonance for the $K^- pp \to \Sigma^0 p$ channel and that of the $\Sigma^0(1385)$ resonance in the case of $K^- pp \to \Lambda p$ \cite{amadeus19}. However, we must note that $\Sigma^0(1385)$ resonance couples to $K^- p$ in a p-wave, making its contribution to antikaon absorption negligible, as was shown in Ref.~\cite{sjPRC12}. According to our model and also Ref.~\cite{sjPRC79}, the most important contribution to the $K^-pp\rightarrow \Sigma^0 p$ channel comes indeed from $\pi^0$ exchange due to the enhanced effect of the $\Lambda(1405)$. On the other hand, the process $K^-pp\rightarrow \Lambda p$ proceeds mainly through the $K^-$ exchange mechanism, dominated again by the $\Lambda(1405)$ and not by the $\Sigma^0(1385)$ resonance. Thus Eq.~(10) in Ref.~\cite{amadeus19} is questionable. 

In Fig.~\ref{fig:ratio_LpiSpi_PB}, we show our result for the ratio $R^*$ as a function of the relative density $\rho/\rho_0$, calculated using the free-space (black) and Pauli blocked (red) amplitudes derived from the BCN (left) and P (right) models. In the region of experimentally relevant densities (gray band), the ratio $R^*$ of mesonic rates to $\Lambda \pi^0$ and $\Sigma^0\pi^0$ final states calculated with Pauli blocked amplitudes turns out to be again smaller than the free-space ratio for both interaction models, being of the order of $0.2$ or lower. This is substantially different from the ratio $R$ of non-mesonic processes to $\Lambda^0 p$ and $\Sigma^0 p$ channels presented in Fig.~\ref{fig:ratio_Lp_S0p_PB}. 
It is to be noted that the ratio $R^*$ diverges for $B_{K^-}=50$~MeV around saturation density since we are approaching $\pi \Sigma$ threshold (see Fig.~\ref{fig:deltaEs}). We have checked that the non-zero value of kaon momentum has again a negligible effect on the ratio $R^*$ in the region of experimentally relevant densities.

\begin{table}[t!]
\caption{Primary-interaction ratios (in $\%$) for mesonic and non-mesonic absorption of $K^-$ in nuclear matter, calculated with free-space and Pauli blocked amplitudes in the BCN model for $B_{K^-}=0$~MeV and $p_{K^-}=0$~MeV/c. The errors denote the uncertainty due to the cut-off dependence. The experimental data corrected for primary interaction are shown for comparison.}
{\footnotesize
 \begin{tabular}{l|c|c|c|c||c|c}\label{tab:primary_ratios}
 BCN & \multicolumn{2}{c|}{$0.3 \rho_0$} & \multicolumn{2}{c||}{$0.5 \rho_0$}  & \multicolumn{2}{c}{Exp.~\cite{bubble3}}  \\ \hline
mesonic ratio & ~free-space~ & ~Pauli~ & ~free-space~ & ~Pauli~ & ~~~~~~$^{4}$He~~~~~~ & $^{12}$C \\ \hline \hline
$\Sigma^+ \pi^- / K^-$ & 19.6 $\pm$ 0.6& 28.8 $\pm$ 0.7& 22.3 $\pm$ 0.9& 28.6 $\pm$ 1.0 & 31.2 $\pm$ 5.0 & 29.4 $\pm$ 1.0\\
$\Sigma^- \pi^0 / K^-$ & 6.2 $\pm$ 0.2 & 5.7 $\pm$ 0.1& 4.3 $\pm$ 0.2 & 5.5 $\pm$ 0.2 & 4.9 $\pm$ 1.3 & 2.6 $\pm$ 0.6\\
$\Sigma^- \pi^+ / K^-$ & 21.9 $\pm$ 0.6 & 14.8 $\pm$ 0.4& 16.5 $\pm$ 0.7 & 9.6 $\pm$ 0.3 & 9.1 $\pm$ 1.6 & 13.1 $\pm$ 0.4\\
$\Sigma^0 \pi^- / K^-$ & 6.2 $\pm$ 0.2 & 5.7 $\pm$ 0.1& 4.4 $\pm$ 0.2 &5.6 $\pm$ 0.2 & 4.9 $\pm$ 1.3 & 2.6 $\pm$ 0.6\\
$\Sigma^0 \pi^0 / K^-$ & 18.1 $\pm$ 0.5 & 19.2 $\pm$ 0.5& 17.6 $\pm$ 0.7 & 16.6 $\pm$ 0.6 & 17.7 $\pm$ 2.9 & 20.0 $\pm$ 0.7\\
$\Lambda \pi^0 / K^-$ & 3.8 $\pm$ 0.1 & 3.5 $\pm$ 0.1 & 2.9 $\pm$ 0.1 & 3.6 $\pm$ 0.1 & 5.2 $\pm$ 1.6 & 3.4 $\pm$ 0.2\\ 
$\Lambda \pi^- / K^-$ & 7.6 $\pm$ 0.2 &7.0 $\pm$ 0.2& 5.8 $\pm$ 0.2 &7.5 $\pm$ 0.3  & 10.5 $\pm$ 3.0 & 6.8 $\pm$ 0.3\\ \hline
total 1N ratio& 83.3 $\pm$ 2.4 & 84.6 $\pm$ 2.2& 73.9 $\pm$ 3.9 &77.1 $\pm$ 2.8 & 83.5 $\pm$ 7.1 & 77.9 $\pm$ 1.6\\ \hline 
$R_{\pm}=\frac{(\Sigma^+ \pi^-)}{(\Sigma^- \pi^+)}$ & 0.9 & 1.9 & 1.4 & 3.0 & 3.5 $\pm$ 1.0 & 2.24 $\pm$ 0.12 \\ 
$R_{pn}=\frac{(\Sigma^+ \pi^-)+(\Sigma^- \pi^+)}{(\Sigma^- \pi^0)}$ & 6.7 & 7.7 & 9.0 & 7.0 & 9.0 $\pm$ 4.0 & 16.3 $\pm$ 4.0  \\ \hline \hline
non-mesonic ratio & ~free-space~ & ~Pauli~ & ~free-space~ & ~Pauli~ & \multicolumn{2}{c}{76\% CF$_3$Br + 24\% C$_3$H$_8$~\cite{bubble1}}  \\ \hline 
$(\Lambda p + \Lambda n +\Sigma^0 p + \Sigma^0 n)/ K^-$ & 8.0 $\pm$ 1.2 & 7.2 $\pm$ 1.1  & 12.4 $\pm$ 1.6 & 10.9 $\pm$ 1.4  &  \multicolumn{2}{c}{14.1 $\pm$ 2.5$~^{\rm{a}}$}  \\
$(\Sigma^- p + \Sigma^- n)/ K^-$  & 5.8 $\pm$ 0.8 & 4.3 $\pm$ 0.6 & 8.1 $\pm$ 0.9 & 5.6 $\pm$ 0.6  & \multicolumn{2}{c}{7.3 $\pm$ 1.3$~^{\rm{a}}$} \\
$\Sigma^+ n / K^-$  & 2.8 $\pm$ 0.4 &  3.8 $\pm$ 0.5 & 5.6 $\pm$ 0.6 & 6.4 $\pm$ 0.7 & \multicolumn{2}{c}{4.3 $\pm$ 1.2$~^{\rm{a}}$}  \\
$(\Sigma^0 p + \Sigma^0 n)/ K^-$  & 3.9 $\pm$ 0.5 & 3.7 $\pm$ 0.5 & 6.2 $\pm$ 0.7 &  5.6 $\pm$ 0.6  & \multicolumn{2}{c}{ - } \\ \hline 
total 2N ratio & 16.7 $\pm$ 2.4 & 15.4 $\pm$ 2.2 & 26.1 $\pm$ 3.0 & 22.9 $\pm$ 2.8  & \multicolumn{2}{c}{25.7 $\pm$ 3.1 \footnote{multi-nucleon capture rate}}  \\

\end{tabular}}
 
\end{table} 

In Table~\ref{tab:primary_ratios} we present primary-interaction branching ratios for mesonic and non-mesonic $K^-$ absorption into different final states, calculated for $p_{K^-}=0$~MeV/c and $B_{K^-}=0$~MeV using the free-space and Pauli blocked BCN amplitudes at $0.3\rho_0$ and $0.5\rho_0$. The branching ratios from bubble chamber experiments \cite{bubble1, bubble3}, corrected for secondary interactions of the primary particles created in the absorbing nucleus, are listed in column `Exp.'. The mesonic branching ratios calculated using the Pauli blocked amplitudes are in better agreement with experimental data than the free-space ratios, especially in the case of $\Sigma^+\pi^-$ and $\Sigma^- \pi^+$ final states. The values of the calculated $\Sigma\pi$ branching ratios shown in Table~\ref{tab:primary_ratios} are consistently related to the behavior of the Pauli blocked amplitudes shown in Fig.~\ref{fig:abs_t}. The magnitude of the Pauli blocked $\Sigma^+ \pi^-$ amplitude around $0.3\rho_0$ is the largest, followed by $\Sigma^0\pi^0$ and finally by $\Sigma^- \pi^+$. The magnitudes of the free-space amplitudes are in different order at this density. Consequently, the ratio $R_{\pm}$ between the $\Sigma^+\pi^-$ and $\Sigma^-\pi^+$ final states gets enhanced by about a factor of two when employing Pauli blocked amplitudes and is in much better agreement with the experiment.
The non-mesonic ratios calculated with Pauli blocked amplitudes presented at the bottom of Table~\ref{tab:primary_ratios} show reasonable agreement with available experimental data as well. It is to be noted that the quoted experimental fractions correspond to global $K^-$ multi-nucleon absorption ratios~\cite{bubble2}, thus including $K^-$ absorption on three and more nucleons which are not considered in our calculations. We assume that the $3N$ and $4N$ absorption processes are less important than the $2N$ absorption ones.

Next, we compare the $K^-NN$ absorption fractions with directly measured values, uncorrected for the effect of secondary interactions, obtained in absorption of low-energy $K^-$ on $^4$He in bubble chamber experiments \cite{katzPRD70}. We should then consider the effect of final state interaction, mainly the $\Sigma N - \Lambda N$ conversion processes, in our results. A proper microscopic quantum-mechanical evaluation of these effects would require the consideration of higher-order diagrams implementing the multiple scattering of $\Lambda N$ and $\Sigma N$ states, coupled by a realistic $YN$ interaction, as well as the possibility of re-scattering and conversion on secondary nucleons. This is out of the scope of the present work. Instead, we implement $\Sigma-\Lambda$ conversion corrections by simply reshuffling strength from channels with $\Sigma^-$, $\Sigma^0$ and $\Sigma^+$ hyperons into channels with a $\Lambda$ baryon, a procedure commonly followed in previous experimental works \cite{bubble3,katzPRD70}. In Table~\ref{tab:corrected_ratios}, we present the two-nucleon absorption fractions calculated for $B_{K^-}=0$~MeV and $p_{K^-}=0$~MeV/c at two different nuclear matter densities within the BCN model. The fractions implement the $\Sigma - \Lambda$ conversion using two different prescriptions~\cite{katzPRD70}. In column a) we consider 60\% for $\Sigma^+ - \Lambda$ conversion, 22.5\% for $\Sigma^- - \Lambda$ conversion, and 72\% for $\Sigma^0 - \Lambda$ conversion. In column b) we take a 50\% conversion probability for all $\Sigma$'s. In the upper half of Table~\ref{tab:corrected_ratios}, we observe that the two-nucleon absorption ratios for the respective channels produce results that, in general, are very close to or in agreement with the experimental data. The total two-nucleon ratio is consistent with the experimental data, especially for the value obtained at $0.3\rho_0$. It is to be noted that the total $2N$ ratio is also compatible with the latest result of the AMADEUS collaboration~\cite{amadeus19} on $K^-$ two-nucleon absorption in carbon, BR$(K^-2NA \rightarrow YN)= (16 \pm 3\text{(stat.)}^{+4}_{-5}\text{(syst.)})\%$. In the lower half of Table~\ref{tab:corrected_ratios}, we present branching ratios for the total $\Sigma^+$, $\Sigma^-$, $\Sigma^0$ and $\Lambda$ production stemming from both $K^-$ single- and two-nucleon absorption. The branching ratios including $\Sigma - \Lambda$ conversion show very good agreement with experimental data. Moreover, the total $\Sigma^+$ and $\Sigma^-$ ratios tend to favor option b) for the $\Sigma - \Lambda$ conversion rate.

\begin{table}[t!]
\caption{Primary-interaction ratios (in $\%$) for non-mesonic and total $K^-$ absorption in matter and corresponding ratios corrected for $\Sigma - \Lambda$ conversion with different conversion rates: a) 60\% for $\Sigma^+ - \Lambda$, 22.5\% for $\Sigma^- - \Lambda$, 72\% for $\Sigma^0 - \Lambda$, b) 50\% for all $\Sigma$'s, calculated with the Pauli blocked BCN amplitudes for $B_{K^-}=0$~MeV and $p_{K^-}=0$~MeV/c. The errors denote the uncertainty due to the cut-off dependence. The experimental data are shown for comparison.} 
{\footnotesize
 \begin{tabular}{l|c|c||c|c|c|c||c}\label{tab:corrected_ratios}
  BCN & ~$0.3 \rho_0$~ & ~$0.5 \rho_0$~ & \multicolumn{2}{c|}{$0.3 \rho_0$ + $\Sigma$-$\Lambda$ conv.} & \multicolumn{2}{c||}{$0.5 \rho_0$ + $\Sigma$-$\Lambda$ conv.} & Exp.~\cite{katzPRD70} \\ \hline
  non-mesonic ratio &  & & ~~~~~a)~~~~~ & b) & ~~~~~a)~~~~~ & b) & $^{4}$He  \\ \hline \hline
$(\Lambda N +\Sigma^0N)/ K^-$ &7.2 $\pm$ 1.1 & 10.9 $\pm$ 1.4& 10.5 $\pm$ 1.5  & 11.3 $\pm$ 1.6 & 16.0 $\pm$ 0.2 & 16.9 $\pm$ 2.1 &  11.7 $\pm$ 2.4  \\
$(\Sigma^- N)/ K^-$  &4.3 $\pm$ 0.6&5.6 $\pm$ 0.6&3.4 $\pm$ 0.4  & 2.2 $\pm$ 0.3 & 4.4 $\pm$ 0.5 & 2.8 $\pm$ 0.3 & 3.6 $\pm$ 0.9 \\
$\Sigma^+ n / K^-$   &3.8 $\pm$ 0.5 &6.4 $\pm$ 0.7  &1.5 $\pm$ 0.2  & 1.9 $\pm$ 0.3 &2.6 $\pm$ 0.3 & 3.2 $\pm$ 0.4 & 1.0 $\pm$ 0.4  \\
$(\Sigma^0 N)/ K^-$  & 3.7 $\pm$ 0.5 &5.6 $\pm$ 0.6 &1.0 $\pm$ 0.1  & 1.9 $\pm$ 0.2 & 1.6 $\pm$ 0.2 & 2.8 $\pm$ 0.3 & 2.3 $\pm$ 1.0  \\ \hline 
total 2N ratio  & 15.4 $\pm$ 2.2 &  22.9 $\pm$ 2.8 &\multicolumn{2}{c|}{15.4 $\pm$ 2.2}  & \multicolumn{2}{c||}{22.9 $\pm$ 2.8} & 16.4 $\pm$ 2.6  \\ \hline \hline
total ratio & & & &  &  & &   \\
$\Sigma^+ / K^-$  & 32.6 $\pm$ 0.2& 35.0 $\pm$ 0.3   &13.0 $\pm$ 0.1  & 16.3 $\pm$ 0.1 &14.0 $\pm$ 0.1 & 17.5 $\pm$ 0.1 & 17.0 $\pm$ 2.7  \\
$\Sigma^- / K^-$ & 24.8 $\pm$ 0.1 & 20.8 $\pm$ 0.1  &19.21 $\pm$ 0.04  & 12.39 $\pm$ 0.03 & 16.1 $\pm$ 0.1 & 10.40 $\pm$ 0.05 & 13.8 $\pm$ 1.8  \\
$\Sigma^0 / K^-$  & 28.7 $\pm$ 0.1 & 27.7 $\pm$ 0.2  &8.03 $\pm$ 0.04  & 14.3 $\pm$ 0.1 & 7.76 $\pm$ 0.05 & 13.9 $\pm$ 0.1 & 10.8 $\pm$ 5.0  \\
$\Lambda / K^-$ & 14.0 $\pm$ 0.3 & 16.5 $\pm$ 0.4  &59.7 $\pm$ 0.1  & 57.0 $\pm$ 0.2 &62.1 $\pm$ 0.1 & 58.2 $\pm$ 0.2 & 58.4 $\pm$ 5.7 \\ \hline
$\Sigma^+/ \Sigma^-$ & 1.31 $\pm$ 0.01 & 1.69 $\pm$ 0.2 & 0.678 $\pm$ 0.006 & 1.31 $\pm$ 0.01 & 0.87 $\pm$ 0.01 & 1.69 $\pm$ 0.2 & 1.2 $\pm$ 0.2 \\
 \end{tabular}}
 
\end{table} 

We are aware that a direct comparison of our nuclear matter results with the $^4$He data has limitations since some of the effects studied here, like Pauli blocking, must be handled in a way that takes into account the finite size of the nucleus.
However, we believe that such comparison is still meaningful as it provides indications that can be useful for guiding future studies performed directly in finite nuclei. In addition, it serves as a consistent check of the calculated ratios, thereby strengthening the validity of our microscopic model.

\section{Conclusions} 
\label{conclusions}

We have developed a microscopic model for the $K^-$ absorption on two nucleons in symmetric nuclear matter. The $K^-$ two-nucleon absorption process has been described within a meson-exchange picture, employing the $K^-N$ scattering amplitudes derived from chiral coupled channel meson-baryon interactions, namely the Prague (P) and Barcelona (BCN) models. Contrary to a similar calculation \cite{sjPRC12}, we have taken into account the Pauli blocking effect in the $K^-N$ amplitudes. 
%\textbf{In the evaluation of the $K^-NN$ potential, we considered the uncertainty tied to the short-range correlations in an effective way.}

We have derived the $K^-$ optical potential as a function of nuclear matter density, including $K^-NN$ (non-mesonic) as well as $K^-N$ (mesonic) absorption processes. Both contributions are substantially affected by the Pauli correlations due to the dominance of the subthreshold $\Lambda(1405)$ resonance in the $K^-N$ amplitudes. The Pauli blocked amplitudes reduce the depth of the absorptive potentials to about one half with respect to the potentials obtained with free-space amplitudes at saturation density. At lower densities, within the $0.15-0.3\rho_0$ region relevant for low-energy antikaon absorption in nuclei, the reduction amounts to 
$25\%$. This result is observed in both interaction models, P and BCN, used in our calculations. Moreover, we have obtained very similar $K^-N$ and $K^-NN$ optical potentials with both chiral models. The absorptive $K^-$ potentials agree with each other up to $0.4\rho_0$. We have also derived the real part of the $K^-NN$ optical potential for the first time. It is mildly repulsive in the whole density region probed. Next, we explored the dependence of the $K^-$ potential on a kaon momentum as well as the uncertainty in the $K^-NN$ potential due to different values of the cut-off parameter used in our approach.

We have calculated single-nucleon and two-nucleon antikaon absorption fractions, as well as several branching ratios, which have been compared with data from old bubble chamber and recent counter experiments. At typical densities for absorption of low energy antikaons, we find $1N$- and $2N$-absorption fractions close to 80\% and 20\%, respectively, in agreement with bubble chamber experiments. The single-nucleon absorption fraction decreases with density (energy) and crosses the two-nucleon fraction at or below the saturation density. The obtained single-nucleon branching ratios into various hyperon-pion channels are in very good agreement with old bubble chamber data. This is only achieved when the Pauli blocked amplitudes are employed, as they substantially enhance the $\Sigma^+ \pi^-$ over the $\Sigma^-\pi^+$ rate. Consequently, we obtain $R_{\pm}=1.9$ at $0.3\rho_0$, which is close to the experimental observations \cite{bubble1,bubble3}. The total two-nucleon absorption fraction amounts to 15\% at  $0.3\rho_0$, which compares very well with $^4$He bubble chamber results \cite{bubble4} as well as with the AMADEUS data~\cite{amadeus19}. After incorporating the effect of $\Sigma-\Lambda$ conversion, our calculated two-nucleon branching ratios into several final states are in good agreement with raw bubble chamber data \cite{bubble4}.

Finally, we have compared the primary $K^-pp$ absorption rates into $\Lambda p$ and $\Sigma^0 p$ final states with the recent measurements of the AMADEUS collaboration for the `quasi-free' production of $\Lambda(\Sigma^0) p$ pairs \cite{amadeus19}. Using the free-space amplitudes, we obtained a value of around 1.1 for the ratio $R=\text{BR}(K^-pp\to\Lambda p)/\text{BR}(K^-pp\to\Sigma^0 p)$, at the edge of the acceptable measured range $0.7\pm0.2(\text{stat.})^{+0.2}_{-0.3}(\text{syst.})$. The Pauli blocking effect reduced the calculated ratio to be close or below 1 for the experimentally accessible densities, well within the experimental errors. This again shows the importance of Pauli correlations in the medium. We have also noticed that the $K^-pp\to\Lambda p$ and $K^-pp\to\Sigma^0 p$ processes are both dominated by the $\Lambda(1405)$ resonance. This, together with the fact that the $KN\Sigma$ Yukawa coupling is relatively small, makes the  $K^-pp\to\Sigma^0 p$ process to be essentially driven by $\pi$-exchange, while  $K$-exchange dominates the $K^-pp\to\Lambda p$ channel. This observation prevents from relating the ratio $R$ to the ratio for mesonic processes $R^*=\text{BR}(K^-p\to\Lambda \pi^0)/\text{BR}(K^-p\to\Sigma^0 \pi^0)$, which in our model amounts to be around 0.2 or smaller at the experimentally relevant densities. 

In summary, the results produced within our microscopic model for $K^-NN$ absorption are in very good agreement with available experimental data. The model seems appropriate to be tested in future applications, such as self-consistent calculations of kaonic atoms and $K^-$-nuclear quasibound states, and the predicted results await further confrontation with new experimental data. 

\section*{Acknowledgments}

A.R. acknowledges support from the Spanish Ministerio de Economia y Competitividad (MINECO) under the project MDM-2014-0369 of ICCUB (Unidad de Excelencia 'Mar\'\i a de Maeztu'), 
and, with additional European FEDER funds, under the contract 
FIS2017-87534-P. J.H. acknowledges support from the GACR grant no. 19-19640S and also financial support from the OP VVV project KINEO, reg. num. CZ.02.2.69/0.0/0.0/16\_027/0008491, which enabled her stay at the University of Barcelona.
We thank A. Ciepl\'{y} for providing us with the Prague model amplitudes. We are grateful to J. Mare\v{s} and E. Friedman for careful reading of the manuscript.

\appendix
\section{$K^-$ single-nucleon optical potential}
\label{appendix:a}
Here, we present details of the derivation of the $K^-N$ optical potential in nuclear matter.
The $K^-N$ self-energy corresponding to Fig.~\ref{fig:one_loop_diagrams} reads
 \begin{align}\label{eq:PiKN1}
  \Pi_{K^-N}(p)= {\rm i}\, |t_{K^-N\rightarrow \pi Y}|^2 \int \frac{d^4 q}{(2\pi)^4} U_{YN}(p-q) \frac{1}{{q}^2-m_{\pi}^2+{\rm i}\eta}~,
 \end{align}
 where $q=(q_0,\vec{q}\,)$ is the pion 4-momentum, $m_{\pi}$ is the pion mass, $p=(p_0,\vec{p}_{K^-}\,)$ is the kaon 4-momentum, $t$ denotes the $K^-N\rightarrow \pi Y~(Y=\Lambda,~\Sigma)$ t-matrix, and $U_{YN}$ is the hyperon-nucleon Lindhard function defined as:
\begin{equation}\label{eq:Uyn}
 U_{YN}(p-q)= \nu \int \frac{d^3 k}{(2 \pi)^3} \frac{m_N}{E_N}\frac{m_Y}{E_Y} \frac{\theta (k_F -| \vec{k}|)}{p_0-q_0 +E_N(\vec{k}) - E_Y(\vec{k}+\vec{p}_{K^-}-\vec{q}\,) + {\rm i}\eta}~.
\end{equation}
Here, $\nu=2$ is the spin degeneracy factor, $E_N (E_Y)$ and $m_{N} (m_Y)$ are the nucleon (hyperon) energy and mass, respectively, $\vec{k}$ is the nucleon momentum and $k_F$ is the nucleon Fermi momentum in nuclear matter of density $\rho$, with $\rho=2k_F^3/3\pi^2$. Taking an average nucleon momentum value $\langle k \rangle=\sqrt{\frac{3}{5}}k_F$ in the energy denominators allows us to perform the $k$-momentum integral as:
\begin{equation*}
 \nu \int \frac{d^3k}{(2 \pi)^3} \theta (k_F -| \vec{k}|) = \frac{\rho}{2} \ , 
\end{equation*}
and integrating over the meson energy $q_0$ in Eq.~\eqref{eq:PiKN1} we obtain
\begin{align}\label{eq:PiKN2}
  \Pi_{K^-N}(p)= &|t_{K^-N\rightarrow \pi Y}|^2~\frac{\rho}{2}  \nonumber \\
  &\times \int \frac{d^3 q}{(2\pi)^3} \frac{1}{2\omega_{\pi}} \frac{m_N}{\langle E_N \rangle}\frac{m_Y}{ E_Y } \frac{1}{E_{K^-} - \omega_{\pi} + \langle E_N \rangle - E_Y(\langle \vec{k}\rangle + \vec{p}_{K^-}-\vec{q}\,) + {\rm i}\eta}~,
\end{align}
with
\begin{equation}\label{eq:EN}
\langle E_N \rangle =  \sqrt{m_N^2 + \frac{3}{5} k_F^2} + V_N\frac{\rho}{\rho_0}~,
\end{equation}
 where we have assumed a nucleon attractive nuclear potential of $V_N=-50$~MeV at saturation density $\rho_0=0.169$~fm$^{-3}$.
 
 By taking the imaginary part of the propagator in Eq.~\eqref{eq:PiKN2}:
\begin{align}
\frac{1}{E_{K^-}-\omega_{\pi}+\langle E_N \rangle- E_Y(\langle \vec{k}\rangle +\vec{p}_{K^-}-\vec{q}\,) +{\rm i}\eta} = & \frac{1}{E_{K^-}-\omega_{\pi} +\langle E_N \rangle- E_Y(\langle \vec{k}\rangle +\vec{p}_{K^-}-\vec{q}\,)} \nonumber \\
& - {\rm i}\, \pi \delta(E_{K^-}-\omega_{\pi}+\langle E_N \rangle- E_Y(\langle \vec{k}\rangle +\vec{p}_{K^-}-\vec{q}\,))~,
\end{align}
i.~e. by putting the one-particle one-hole (1p-1h) excitation on-shell, and integrating over $\vec{q}$, we obtain the imaginary part of the $K^-N$ optical potential 
\begin{equation}\label{eq:imVknpiY}
\text{Im}V_{K^-N\rightarrow \pi Y}(p) = \frac{\text{Im} \Pi_{K^-N}}{2 E_{K^-}}= -\frac{1}{2 E_{K^-}}\frac{1}{4\pi} \frac{\rho}{2}~|t_{K^-N\rightarrow \pi Y}|^2 \frac{\overline{q}}{\langle E_N \rangle} \frac{m_N m_Y}{E_Y(\overline{q}) + \omega_{\pi}(\overline{q})}~,
\end{equation}
where we have considered a kaon with energy
\begin{equation}\label{eq:EK}
 E_{K^-}=m_{K^-}-B_{K^-}\frac{\rho}{\rho_0}~, 
\end{equation}
where $m_{K^-}$ and $B_{K^-}$ are the kaon mass and binding energy at $\rho_0$, respectively. The quantity $\overline{q}$ denotes the on-shell pion momentum stemming from the energy conservation
\begin{equation}
 \overline{q}\,^2 = \frac{\left(s_{KN}-(m_Y^{\prime}+m_{\pi})^2\right)\left(s_{KN}-(m_Y^{\prime}-m_{\pi})^2\right)}{4s_{KN}}~,
\end{equation}
where $s_{KN}=(E_{K^-}+\langle E_N \rangle)^2$, $m_Y^{\prime\, 2}=m_Y^2 + \frac{3}{5}k_F^2 + p^2_{K^-}$, and $E_Y(\overline{q})=\sqrt{m_Y^{\prime~2} + \overline{q}\,^{2}}$. 

\section{$K^-$ two-nucleon optical potential}
\label{appendix:b}
In this Appendix, we give a detailed description of derivation of the $K^-NN$ optical potential in nuclear matter.
The $K^-NN$ self-energy corresponding to the 2FL diagrams (a) and (b) shown in Fig.~\ref{fig:direct_diagrams} is given by:
\begin{align}\label{eq:PiB1B2_direct}
  \Pi_{K^-NN}^{\rm 2FL}(p)=& -{\rm i} t_{B_1 x} t^*_{B_1 x} V_{B_2 N_2 x} V_{B_2 N_2 x} \int \frac{d^4 q}{(2\pi)^4} U_{B_1N_1}(p-q) U_{B_2N_2}(q) \notag \\ & \times (-\vec{q}^{\,2}) \frac{1}{q^2-m_{x}^2+{\rm i}\eta}~ \frac{1}{q^2-m_{x}^2+{\rm i}\eta}~,
 \end{align}
where $x$ denotes the intermediate exchanged meson with mass $m_{x}$, which can be a kaon, pion or eta meson. The final baryon attached to the incoming kaon vertex, denoted by $B_{1}$ is either $N$ or $Y$ and, correspondingly, the baryon $B_2$ emitted from the other vertex can be $Y$ or $N$. The two-body t-matrix for the $K^-N_1\rightarrow B_1 x$ channel is denoted by $t_{B_1 x}$, the strength of the Yukawa p-wave type meson-baryon-baryon coupling vertices is given by  $V_{B_2 N_2 x}= \alpha \displaystyle\frac{D+F}{2f_{\pi}} + \beta \displaystyle\frac{D-F}{2f_{\pi}}$, with $D+F= 1.26$, $D-F=0.33$, $f_{\pi}=93$~MeV, and $\alpha,~\beta$ being SU(3) Clebsch-Gordan coefficients. We note that the trace over spins gives a factor $(-4\vec{q}^{\,2})$, employing a non-relativistic approximation for the Yukawa vertices. However, the factor 4 does not appear explicitly in Eq.~\eqref{eq:PiB1B2_direct} because it is implicitly generated by the spin-degeneracy factors $\nu$ in the hyperon-nucleon and nucleon-nucleon 
Lindhard functions, the former introduced in Eq.~~\eqref{eq:Uyn} and the later given by
\begin{align}\label{eq:Unn}
 U_{NN}(p-q)=& \nu \int \frac{d^3 j}{(2 \pi)^3} \frac{m_N^2}{E_N^2} \left[  \frac{\theta (k_F -|\vec{j}|) \theta (|\vec{j}+\vec{p}_{K^-}-\vec{q}\,|-k_F)}{p_0-q_0+E_N(\vec{j})- E_N(\vec{j}+\vec{p}_{K^-} - \vec{q}\,)+ {\rm i}\eta} \right. \\[1ex] \notag & + \left.  \frac{\theta(|\vec{j}|-k_F) \theta (k_F -|\vec{j}+\vec{p}_{K^-}-\vec{q}\,|)}{-p_0+q_0-E_N(\vec{j})+ E_N(\vec{j}+\vec{p}_{K^-} - \vec{q}\,)+ {\rm i}\eta}\right] ~.
\end{align}
The combination of Lindhard functions contributing to the 2FL diagram (a) in Fig.~\ref{fig:direct_diagrams} is $U_{NN}(p-q) U_{YN}(q)$ while that for diagram (b) is $U_{NN}(q) U_{YN}(p-q)$. We omit the second term in the square brackets in Eq.~\eqref{eq:Unn}, the so-called crossed contribution, because it is much smaller than that of the first term. Further, we apply the same approximations as in Eq.~\eqref{eq:PiKN2}. Then the expression for the $K^-NN$ self-energy coming from either the 2FL(a) or the 2FL(b) diagrams is of the form
\begin{align} \label{eq:PiNN}
 \Pi_{K^-NN}^{\rm 2FL}(p)=& t_{B_1 x} t^*_{B_1 x}  V_{B_2 N_2 x} V_{B_2 N_2 x} \frac{\rho^2}{4} \nonumber \\
 &\times \int \frac{d^3 q}{(2\pi)^3} \frac{\vec{q}^{\,2}\, \theta (|\langle \vec{j} \rangle +\vec{p}_{K^-}-\vec{q}\,|-k_F)}{p_0+2\langle E_N \rangle- E_{B_1}(\langle \vec{j} \rangle + \vec{p}_{K^-} - \vec{q}\,)-E_{B_2}(\langle \vec{k} \rangle + \vec{q}\,) +{\rm i}\eta} \notag \\
 %[1ex] \notag 
 & \times \frac{1}{q_0^2-\vec{q}^{\,2}-m_{x}^2 +{\rm i}\eta} ~ \frac{1}{q_0^2-\vec{q}^{\,2}-m_{x}^2 +{\rm i}\eta} \left(\frac{m_N}{\langle E_N \rangle}\right)^2 \frac{m_{B_1}}{E_{B_1}} \frac{m_{B_2}}{E_{B_2}}~,
\end{align}
where $q_0 = p_0 + \langle E_N \rangle -E_{B_1}$ and $\langle \vec{j} \rangle=\langle \vec{k} \rangle=\sqrt{\frac{3}{5}}k_F$ are the averaged nucleon momenta.

The imaginary part of the $2N$-absorption potential is obtained by putting the 2p-2h excitation on-shell, i.e. retaining the imaginary part of the baryon propagator in Eq.~(\ref{eq:PiNN})
\begin{align}
& \frac{1}{p_0 +2\langle E_N \rangle- E_{B_1}(\langle \vec{j} \rangle + \vec{p}_{K^-} - \vec{q}\,)-E_{B_2}(\langle \vec{k} \rangle +\vec{q}\,) +{\rm i}\eta} =\nonumber \\
&~~~~~~~~~~~~~~~~~~ \frac{1}{p_0+2\langle E_N \rangle- E_{B_1}(\langle \vec{j} \rangle + \vec{p}_{K^-} - \vec{q}\,)-E_{B_2}(\langle \vec{k} \rangle +\vec{q}\,)} \notag \\ &~~~~~~~~~~~~~~~~~ 
- {\rm i} \pi \delta(p_0 +2\langle E_N \rangle- E_{B_1}(\langle \vec{j} \rangle + \vec{p}_{K^-} - \vec{q}\,)-E_{B_2}(\langle \vec{k} \rangle +\vec{q}\,))~.
\end{align}
Next, we set $p_0\to E_{K^-}$, neglect the angular dependence in the energy denominators, and perform an angle average of the Pauli function $\theta (|\langle \vec{j} \rangle + \vec{p}_{K^-} -\vec{q}\,|-k_F)$\footnote{We checked that this approximation is reasonable for low-energy kaons with momentum values up to $\sim 150$~MeV/c.}: 
\begin{align}
  w(\vec{q}\,) = \left\{ \begin{array}{cl}
  0 & \text{if } |\vec{q}\,|< k_F- |\langle \vec{j} \rangle + \vec{p}_{K^-} | \\
  \frac{(|\vec{q}\,|+|\langle \vec{j} \rangle + \vec{p}_{K^-} |)^2 - k_F^2}{4 |\langle \vec{j} \rangle + \vec{p}_{K^-} | |\vec{q}\,|} & \text{if } k_F - |\langle \vec{j} \rangle + \vec{p}_{K^-} | <|\vec{q}\,|< k_F+ |\langle \vec{j} \rangle + \vec{p}_{K^-} | \\
  1 & \text{if } |\vec{q}\,| > k_F + |\langle \vec{j} \rangle + \vec{p}_{K^-} |~,
  \end{array}
  \right.
\end{align}
where the quantity  $|\langle \vec{j} \rangle + \vec{p}_{K^-} |$ in the above expression is to be replaced by the corresponding angle-averaged one $ \sqrt{ \langle j \rangle^2 + p_{K^-}^2}$.
Finally, we obtain the following analytic expression for the imaginary part of the $K^-NN$ optical potential 
\begin{align}
 \text{Im}V_{K^-NN}^{\rm 2FL}(p)= \frac{\text{Im} \Pi_{K^-NN}^{\rm 2FL}}{2 E_{K^-}} =& -\frac{1}{2 E_{K^-}}  t_{B_1 x} t^*_{B_1 x} 
 V_{B_2 N_2 x} V_{B_2 N_2 x} \frac{1}{2 \pi} \frac{\rho^2}{4} \nonumber \\
 & \times \frac{m_{B_1} m_{B_2}}{E_{B_1}(\overline{q})+E_{B_2}(\overline{q})} \left(\frac{m_N}{\langle E_N \rangle}\right)^2 \overline{q} \notag \\[1ex] & \times \overline{q}^{\,2} w(\overline{q}) F^2_H(\overline{q})~ \frac{1}{q_0^2-\overline{q}^{2}-m_{x}^2}~ \frac{1}{q_0^2-\overline{q}^{2}-m_{x}^2}~,
\label{imV2FL}
\end{align}
where $E_{B_{1(2)}}(\overline{q})=\sqrt{m_{B_{1(2)}}^{\prime~2}+\overline{q}^2}$ with $m_{B_{1}}^{\prime~2}=m_{B_{1}}^{~2} + \frac{3}{5}k_F^2+p^2_{K^-}$, $m_{B_{2}}^{\prime~2}=m_{B_{2}}^{~2} + \frac{3}{5}k_F^2$, and 
 $\overline{q}$ is the meson center-of-mass momentum
 \begin{equation}
 \overline{q}^2 = \frac{(s_{K2N}-(m_{B_1}^{\prime}+m_{B_2}^{\prime})^2)(s_{K2N}-(m_{B_1}^{\prime}-m_{B_2}^{\prime})^2)}{4s_{K2N}}~,
\end{equation}
with $s_{K2N}=(E_{K^-}+2\langle E_N \rangle)^2$.
Note that we have incorporated a form-factor 
\begin{equation}\label{eq:form_factor}
F_H(\overline{q})=\displaystyle\frac{\Lambda_C^2}{\Lambda_C^2+\overline{q}^{\,2}}
\end{equation}
in each of the Yukawa vertices of the $K^-NN$ self-energy diagrams, with a cut-off parameter $\Lambda_C=1200$~MeV, a value which is in line with those employed in the meson-exchange models of the $NN$ (Bonn) and $YN$ (J\"{u}lich) interactions \cite{Machleidt:1987hj,Holzenkamp:1989tq}. Neglecting the angular dependence in the energy denominators of Eq.~\eqref{eq:PiNN}, the real part acquires the following form
\begin{align}\label{eq:RePiB1B2}
 \text{Re}V_{K^-NN}^{\rm 2FL}(p) =  \frac{\text{Re}\Pi_{K^-NN}^{\rm 2FL}}{2E_{K^-}} =&
 \frac{1}{2 E_{K^-}} t_{B_1 x} t^*_{B_1 x}  V_{B_2 N_2 x} V_{B_2 N_2 x} \frac{\rho^2}{4} \nonumber \\
& \times
 \int \frac{\vec{q}^{\,2} dq}{2 \pi^2} \left(\frac{m_N}{\langle E_N \rangle}\right)^2 \frac{m_{B_1}}{E_{B_1}} \frac{m_{B_2}}{E_{B_2}}~ \vec{q}^{\,2}  w(\vec{q}\,) F^2_H(\vec{q}\,)~ \nonumber \\[1ex] & \times \frac{1}{E_{K^-}+2\langle E_N \rangle- E_{B_1}(\langle \vec{j} \rangle + \vec{p}_{K^-}-\vec{q}\,)-E_{B_2}(\langle \vec{k} \rangle+\vec{q}\,)}  \notag \\[1ex] &  \times \frac{1}{q_0^2-\vec{q}^{\,2}-m_{x}^2}~ \frac{1}{q_0^2-\vec{q}^{\,2}-m_{x}^2}~.
% \label{reV2FL}
\end{align}

It is to be noted that the contribution to the $K^-NN$ self-energy coming from the 2FL diagrams (c) and (d) of Fig.~\ref{fig:direct_diagrams} is zero due to the null trace over spins, which is related to the fact that there is only one operator
$\vec{\sigma}$ in each fermionic loop.

On the other hand, there are non-negligible contributions to the $K^-NN$ self-energy coming from 1FL diagrams displayed in Fig.~\ref{fig:crossed_diagrams}. The self-energy corresponding to the 1FLA-type diagrams (a) and (b) is of the form
\begin{align}\label{eq:PiB1B2_c2}
 \Pi_{K^-NN}^{\rm 1FLA}(p)= & - {\rm i} t_{B_1 x_1} t^*_{B_2 x_2} V_{B_2 N_2 x_1} V_{B_1 N_2 x_2} \int \frac{d^4 q}{(2\pi)^4} (+2\vec{q}^{~2}) \int \frac{d^4 j}{(2\pi)^4} G_{N_1}(j) G_{B_1}(j+p -q) \notag
 \\[1ex] &\times \int \frac{d^4 k}{(2 \pi)^4} G_{N_2}( k) G_{B_2}(q+k) \frac{1}{q^2-m_{x_1}^2+{\rm i}\eta}~ \frac{1}{q^{\prime\,2}-m_{x_2}^2+{\rm i}\eta}~,
\end{align}
where $G$ denotes the in-medium baryon propagator and $q^\prime=j+p-q-k$. For these diagrams, the trace over spins gives the factor $(+2\vec{q}^{\,2})$. This is because we have assumed to have small values of the kaon and nucleon momenta  ($\vec{p}_{K^-},\vec{j},\vec{k}\sim 0$) in the upper meson-exchange line and the remaining vector, $-\vec{q}$, is opposite in sign than that in the upper meson-exchange line of diagrams 2FL(a) and (b). Taking average values for the $j$ and $k$ dependencies in the propagator of meson $x_2$ allows us to approximate the integrals over these two four-momenta by a factor $(-{\rm i})$ times the $YN$ and $NN$ Lindhard functions, given by Eqs.~\eqref{eq:Uyn} and \eqref{eq:Unn}, respectively.  Again, a global factor of 4 will be removed explicitly from Eq.~\eqref{eq:PiB1B2_c2} because it is already taken into account by the spin degeneracy factors $\nu$ in the Lindhard functions. After employing the same procedure as for the 2FL diagrams we arrive at the following expression for the imaginary $K^-NN$ self-energy for 1FLA-type diagrams:
\begin{align}\label{eq:ImPiB1B2_c2}
\text{Im}V_{K^-NN}^{\rm 1FLA}(p) = \frac{\text{Im} \Pi_{K^-NN}^{\rm 1FLA}}{2 E_{K^-}} =& - \frac{1}{2 E_{K^-}}\frac{1}{2}  t_{B_1 x_1} t^*_{B_2 x_2} V_{B_2 N_2 x_1} V_{B_1 N_2 x_2} \frac{1}{2 \pi} \frac{\rho^2}{4} \nonumber \\
& \times \overline{q}~ \frac{m_{B_1} m_{B_2}}{E_{B_1}(\overline{q})+E_{B_2}(\overline{q})} \left(\frac{m_N}{\langle E_N \rangle}\right)^2  \overline{q}^{\,2} w(\overline{q})~ \notag \\[1ex] & \times \frac{F_H(\overline{q})}{q_0^2-\overline{q}^{\,2}-m_{x_1}^2}~ \frac{F_H(\langle j \rangle+\vec{p}_{K^-}-\overline{q}-\langle k\rangle)}{q_0^{\prime\,2}-\overline{q}^{\,2}-\langle j \rangle^2-\langle k\rangle^2-p^2_{K^-}-m_{x_2}^2}~,
 %\label{imV1FL}
\end{align}
where $q_0^{\prime}=E_{B_1}(\overline{q})-\langle E_N \rangle$. The real part is of the following form
\begin{align}\label{eq:RePiB1B2_c2}
 \text{Re}V_{K^-NN}^{\rm 1FLA}(p) =  \frac{\text{Re}\Pi_{K^-NN}^{\rm 1FLA}}{2 E_{K^-}} =& \frac{1}{2 E_{K^-}}\frac{1}{2} t_{B_1 x_1} t^*_{B_2 x_2} V_{B_2 N_2 x_1} V_{B_1 N_2 x_2} \frac{\rho^2}{4} \nonumber \\
 & \times \int \frac{\vec{q}^{\,2} dq}{2 \pi^2} \left(\frac{m_N}{\langle E_N \rangle}\right)^2 \frac{m_{B_1}}{E_{B_1}} \frac{m_{B_2}}{E_{B_2}}~\vec{q}^{\,2}  w(\vec{q}\,) \notag \\[1ex] &\times \frac{1}{E_{K^-}+2\langle E_N \rangle- E_{B_1}(\langle \vec{j} \rangle+\vec{p}_{K^-}-\vec{q}\,)-E_{B_2}(\langle \vec{k} \rangle+\vec{q}\,)}\notag \\[1ex] & \times \frac{F_H(\vec{q}\,)}{q_0^2-\vec{q}^{\,2}-m_{x_1}^2}~  \frac{F_H(\langle j \rangle+\vec{p}_{K^-}-\vec{q}-\langle k\rangle)}{q_0^{\prime\,2}-\vec{q}^{\,2}-\langle j\rangle^2-\langle k\rangle^2-p^2_{K^-}-m_{x_2}^2}~.
  %\label{reV1FL}
\end{align}

The expression for the self-energy corresponding to the 1FLB-type diagrams (c) and (d) in Fig.~\ref{fig:crossed_diagrams} is very similar to that for the 1FLA-type diagrams (a) and (b), with some differences. First, the four-momentum $q^\prime$ in the upper meson exchange line is $j+p-q-k$ and it is now directed towards a Yukawa-type vertex. This means that the trace over spins yields now a factor $(-2\vec{q}^{~2})$, introducing an overall relative minus sign. There is also a change in the value of $q_0^{\prime}$ which becomes $q_0^{\prime}=\langle E_N \rangle-E_{B_2}(\langle k \rangle+\vec{q}\,)$. Finally, both exchanged mesons in diagrams (c) and (d) are now the same one, hence the vertices in Eqs.~(\ref{eq:ImPiB1B2_c2}) and (\ref{eq:RePiB1B2_c2})
should be replaced by $t_{B_1 x} t^*_{B_1 x} V_{B_2 N_2 x} V_{B_2 N_2 x}$, with $x$ being either a kaon, a pion or an eta meson. 
Notice that, due to the spin traces, the 1FL diagrams have acquired an additional factor 1/2 with respect to the 2FL ones. Moreover, the sign of the 1FLA (1FLB) diagrams is the same (opposite) as that of the 2FL ones.

\end{document}